\newcommand{\myPaperFontSize}{9pt}
\documentclass[a4paper, \myPaperFontSize, twocolumn, fleqn]{extarticle}

\usepackage{graphicx}
\usepackage{amssymb}
\usepackage{lineno}

\graphicspath{{}
{./figures/}
}

\usepackage[american]{babel}
\usepackage[autostyle]{csquotes}
\usepackage[comma,authoryear,sort&compress]{natbib}		
\usepackage{amsmath}					
\usepackage[tight,nice]{units}			
\usepackage{subfig}						

\usepackage{hyperref}		
\usepackage{xcolor}
\definecolor{dark-red}{rgb}{0.45,0.15,0.15}
\definecolor{dark-blue}{rgb}{0.15,0.15,0.4}
\definecolor{medium-blue}{rgb}{0,0,0.5}
\hypersetup{
    colorlinks, 				
    linkcolor={dark-red},		
    citecolor={dark-blue},		
    urlcolor={medium-blue},		
}

\usepackage[labelfont=bf,labelsep=period, font=small]{caption}

\RequirePackage[l1tabu, orthodox]{nag}			
\usepackage{tabularx}							
\usepackage{booktabs}							
\usepackage[affil-it]{authblk}					
\usepackage{soul}								

\usepackage{setspace}							
\usepackage{afterpage}
\usepackage[color]{changebar}							
\cbcolor{yellow}
\usepackage{chngcntr}							

\renewcommand{\topfraction}{0.8}				

\let\oldtabularx\tabularx
\renewcommand{\tabularx}{\small\oldtabularx}

\renewcommand*{\div}{\mathop{\mathrm{div}}\nolimits}
\DeclareMathOperator{\e}{e}
\newcommand*{\fm}[1]{(\ref{#1})}

\newcommand*{\partialFrac}[2]{{\frac{\partial{#1}}{\partial #2}}}
\newcommand{\paragraphBold}[1]{\paragraph{\textbf{#1}}}
\newcommand{\protectedSubref}[1]{\protect\subref{#1}}
\newcommand{\registeredTrademark}{{\textsuperscript{\textregistered}}}
\newcommand{\ASmallVilli}{{A_\mathrm{sm.v}}}
\newcommand{\DP}{{D_p}}
\newcommand{\cBnd}{{c_\mathrm{bnd}}}
\newcommand{\cMax}{{c_\mathrm{max}}}

\newcommand{\cPl}{{c_\mathrm{pl}}}
\newcommand{\cPlAverage}{{\bar{c}_\mathrm{pl}}}
\newcommand{\cTot}{{c_\mathrm{tot}}}

\newcommand{\FMax}{{F_\mathrm{max}}}

\newcommand{\jCTot}{{\vec j_\cTot}}

\newcommand{\kHenry}{{k_\mathrm{hn}}}
\newcommand{\kHill}{{k_\mathrm{hl}}}

\newcommand{\nCot}{{n_\mathrm{cot}}}

\newcommand{\nMax}{{n_\mathrm{max}}}
\newcommand{\nMF}{{n}}

\newcommand{\pFet}{{p_\mathrm{fet}}}
\newcommand{\pMat}{{p_\mathrm{mat}}}
\newcommand{\pONew}{{p_\mathrm{O_2}}}
\newcommand{\PSmallVilli}{{P_\mathrm{sm.v}}}

\newcommand{\RNum}{{R_\mathrm{num}}}
\newcommand{\RNumSquared}{{R_\mathrm{num}^2}}

\newcommand{\SM}{{S_\mathrm{IVS}}}

\newcommand{\SSmallVilli}{{S_\mathrm{sm.v}}}
\newcommand{\sTot}{{S_\mathrm{{tot}}}}
\newcommand{\tObs}{{t_\mathrm{{obs}}}}
\newcommand{\vDye}{{V_\mathrm{dye}}}
\newcommand{\vIVS}{{V_\mathrm{IVS}}}
\newcommand{\vMolar}{{V_\mathrm{m}}}
\newcommand{\vOxygen}{{V_\mathrm{O_2}}}
\newcommand{\vSmallVilli}{{V_\mathrm{sm.v}}}
\newcommand{\vVil}{{V_\mathrm{vil}}}
\newcommand{\bSixty}{\beta_{60}}
\newcommand{\nuOxygen}{\nu_\mathrm{O_2}}
\newcommand{\phiC}{{\phi_\mathrm{c}}}

\newcommand{\rhoBl}{{\rho_\mathrm{bl}}}
\newcommand{\tauD}{{\tau_\mathrm{D}}}
\newcommand{\tauHb}{{\tau_\mathrm{Hb}}}
\newcommand{\tauP}{{\tauHb}}
\newcommand{\tauPas}{{\tau_\mathrm{tr}}}

\title{\huge \bfseries Optimal villi density for maximal oxygen uptake in the human placenta
}

\author[a]{A.S.~Serov
\thanks{Corresponding author. 
\\
E-mail: \texttt{alexander.serov@polytechnique.edu}. 
\\
Corresponding address: Laboratoire de Physique de la Mati\`ere Condens\'ee, Ecole Polytechnique, CNRS, 91128 Palaiseau Cedex, France. Tel.: +33 1 69 33 47 07, Fax: +33 1 69 33 47 99 }
}
\author[b]{C.M.~Salafia}
\author[c,d]{P.~Brownbill}
\author[a]{D.S.~Grebenkov}
\author[a]{M.~Filoche}

\affil[a]{Laboratoire de Physique de la Mati\`ere Condens\'ee, Ecole Polytechnique, CNRS, 91128 Palaiseau Cedex, France}
\affil[b]{Placental Analytics LLC, 93 Colonial Avenue, Larchmont, New York 10538, USA}
\affil[c]{Maternal and Fetal Health Research Centre, Institute of Human Development, University of Manchester, Oxford Road, Manchester M13 9WL, UK}
\affil[d]{St. Mary's Hospital, Central Manchester University Hospitals NHS Foundation Trust, Manchester Academic Health Science Centre, Manchester M13 9WL, UK}

\date{
\today
}

\begin{document}

\maketitle



\begin{abstract}

We present a stream-tube model of oxygen exchange inside a human placenta functional unit~(a placentone). 
The effect of villi density on oxygen transfer efficiency is assessed by numerically solving the diffusion-convection equation in a 2D+1D geometry for a wide range of villi densities. For each set of physiological parameters, we observe the existence of an optimal villi density providing a maximal oxygen uptake as a trade-off between the incoming oxygen flow and the absorbing villus surface. %
The predicted optimal villi density~$0.47\pm0.06$ is compatible to previous experimental measurements. %
Several other ways to experimentally validate the model are also proposed.
The proposed stream-tube model can serve as a basis for analyzing the efficiency of human placentas, detecting possible pathologies and diagnosing placental health risks for newborns by using routine histology sections collected after birth.

\vspace{20pt}


\vspace{20pt}

This article was published in the \emph{Journal of Theoretical Biology} \citep{Serov2015Optimality} and can be accessed by its doi: \href{http://dx.doi.org/10.1016/j.jtbi.2014.09.022}{10.1016/j.jtbi.2014.09.022}.

\end{abstract}


\section{Introduction}
The human placenta is the sole organ of exchange between the mother and the growing fetus.
It consists of two principal components: (i) fetal chorionic villi, and (ii) the intervillous blood basin in which maternal blood flows~(Fig.~\ref{fig:PlacentaScheme}).
The placenta fetal villi have a tree-like structure and are immersed in the maternal blood basin anchoring to the basal plate and supporting the placenta shape~\citep{benirschke_pathology}. Fetal arterial blood goes from the umbilical arteries to the capillaries inside the villi and returns to the umbilical vein, while maternal blood percolates outside this arboreous structure. The exchange of oxygen and nutrients takes place at the surface of the villous tree. 
We aim to understand how the geometry of the villous tree affects the placenta exchange function.

\newcommand{\FirstFigureCaption}{
Structure of the human placenta~\citep[reproduced from][]{Gray1918}.
}
\begin{figure}[htb]  
\center
\includegraphics[width=3 in,natwidth=1280,natheight=1049]
{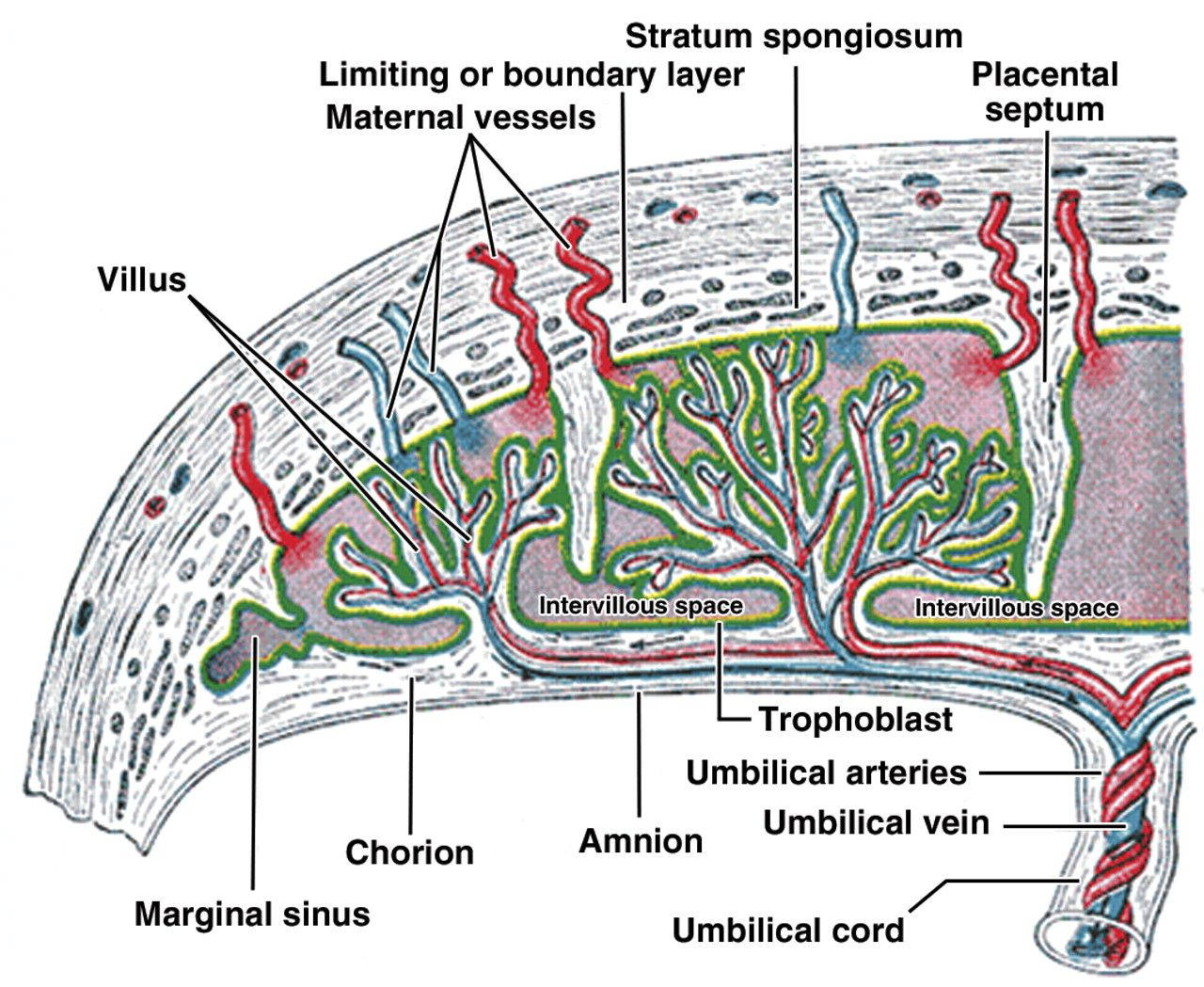}
\caption{\FirstFigureCaption
}
\label{fig:PlacentaScheme}
\end{figure}


The normal development of the baby largely depends on the ability of the placenta to efficiently transfer oxygen, nutrients and other substances. 
Both placenta weight and placenta-fetus weight ratio are associated with the newborns health~\citep{Hutcheon2012,Teng2012}. Moreover, \enquote{fetal origins of adult health} research has identified links between placenta size and risk of heart disease in adults~\citep{Barker1995}. 

Routinely stained 2D histology sections of the placenta collected after birth~(Fig.~\ref{fig:HealthyAndAbnormalPlacentaSections}) may provide valuable insights into its function during pregnancy. 
Figure~\ref{fig:HealthyAndAbnormalPlacentaSections} illustrates our basic idea: 
random sampling of a normal placenta~(Fig.~\ref{fig:HealthyPlacenta}) contains intervillous space~(IVS) and villous tree sections in \enquote{normal} proportions~(which allow efficient function), whereas pre-eclamptic~(Fig.~\ref{fig:PreeclampticPlacenta}, disproportionally large IVS, rare villi) and diabetic~(Fig.~\ref{fig:DiabeticPlacenta}, denser and larger villi) cases exhibit a very different geometry. 

\newcommand{\SecondFigureCaption}{
Typical 2D placenta cross-sections:
\protectedSubref{fig:PreeclampticPlacenta}
pre-eclamptic placenta~(rarefied villi, reduced exchange surface); 
\protectedSubref{fig:HealthyPlacenta} 
normal placenta; 
\protectedSubref{fig:DiabeticPlacenta}
diabetic placenta~(dense villi, reduced surface accessibility). White space is intervillous space~(IVS), normally filled with maternal blood, which has been washed away during the preparation of the slides (some residual red blood cells are still present). The dark shapes are cross-sections of fetal villi. %
The sections are H\&E stained and have been taken in the direction from the basal~(maternal) to the chorionic~(fetal) plate.%
}
\captionsetup[subfigure]{justification=centering}
\begin{figure*}[t]  
\vspace{-4ex}
\centering \subfloat[]{
\includegraphics[width=0.3\linewidth,natwidth=1439,natheight=1050]{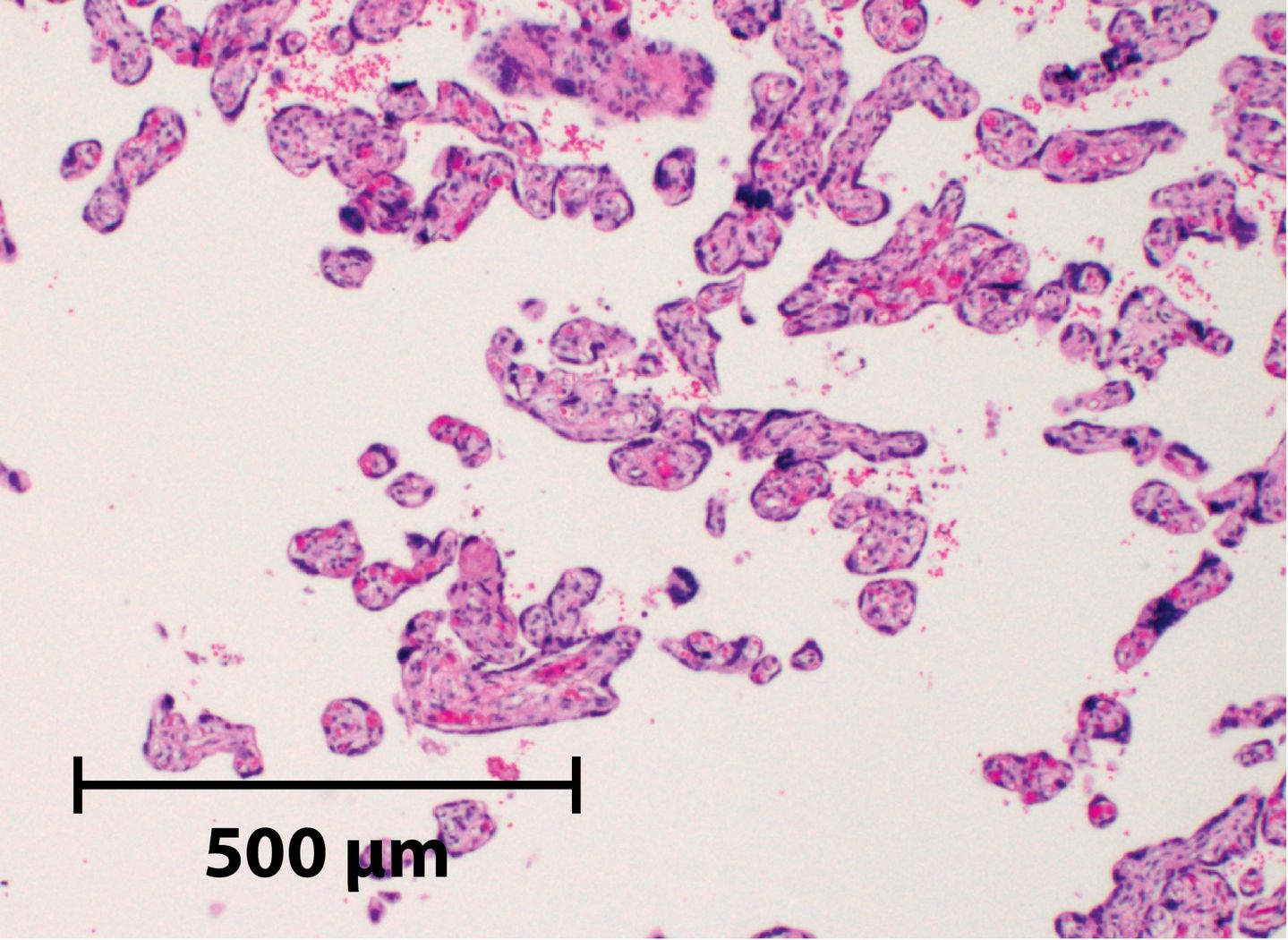}
\label{fig:PreeclampticPlacenta}
}  
\hspace{1ex}
\subfloat[]{
\includegraphics[width=0.3\linewidth,natwidth=1439,natheight=1050]{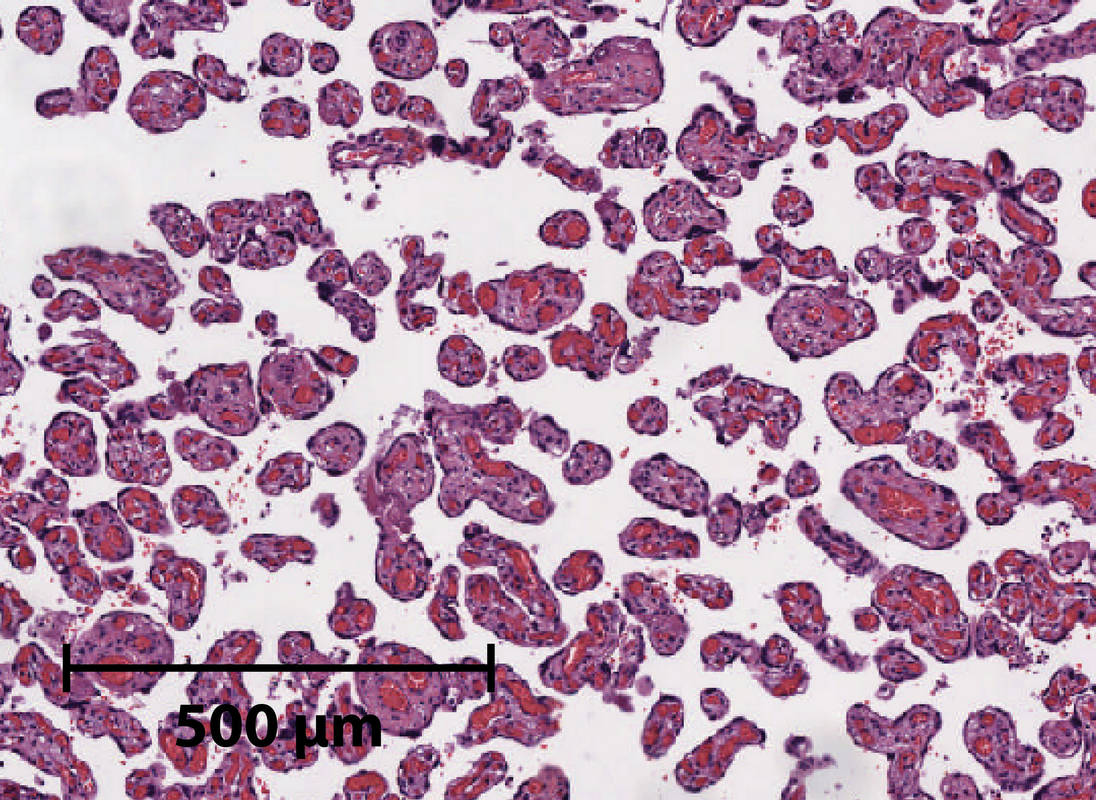}
\label{fig:HealthyPlacenta} }
\hspace{1ex}
\subfloat[]{
\includegraphics[width=0.3\linewidth,natwidth=1439,natheight=1050]{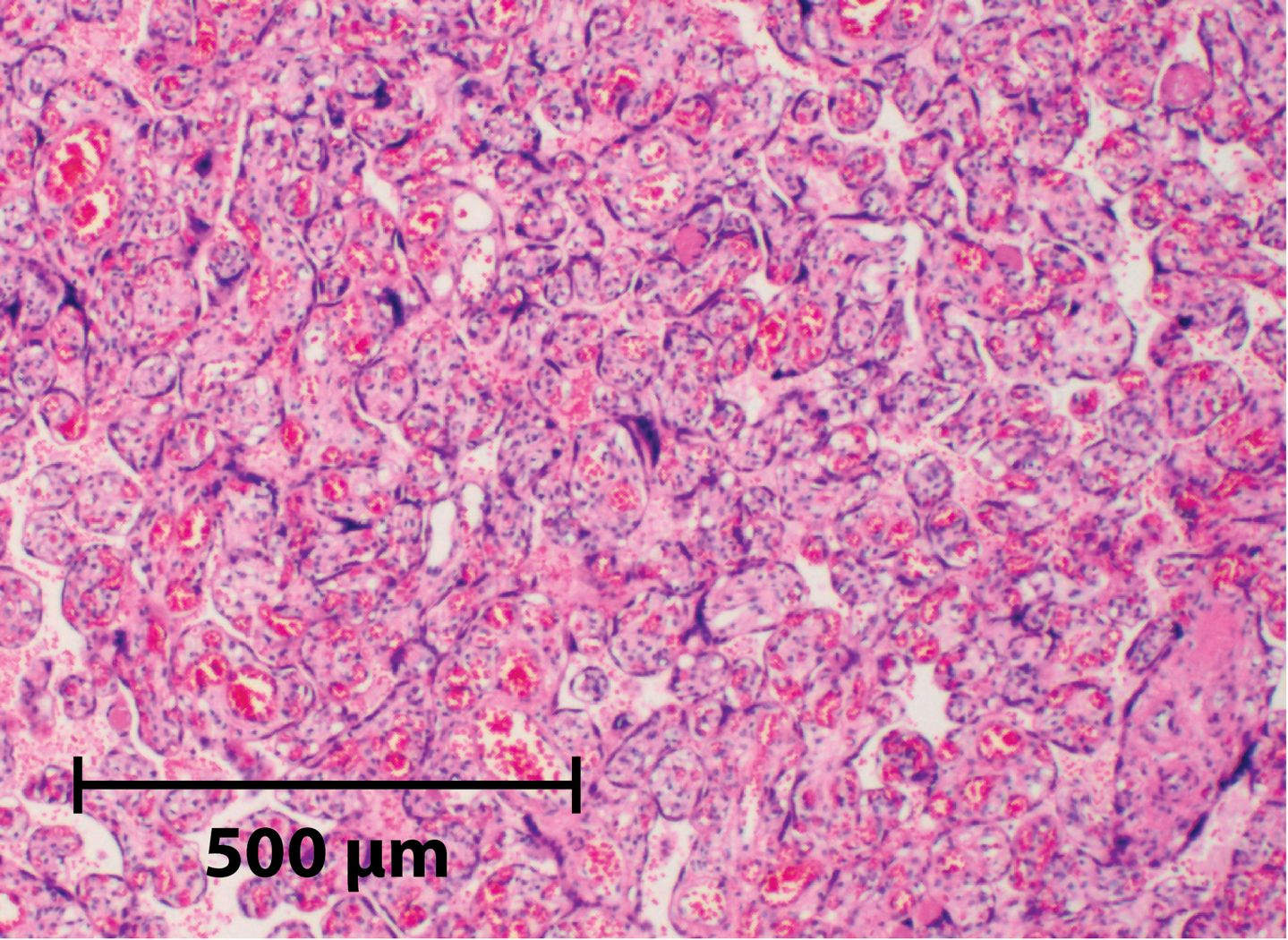}
\label{fig:DiabeticPlacenta} }  
\caption{\SecondFigureCaption
}
\label{fig:HealthyAndAbnormalPlacentaSections}
\end{figure*}


Assessing the relation between the placenta structure and its function is problematic because (i) it is unethical to manipulate the human placenta \emph{in vivo}; (ii) histological and physiological features commonly co-vary~(e.g. changes in villi distribution alter the blood flow); (iii) quantitative histological analysis of the placenta structure can only be performed post-partum when maternal blood flow~(MBF) and fetal blood flow have ceased.
Mathematical modeling and numerical simulations are in this case valuable tools to gain deeper insights into the \emph{in vivo} functioning of the placenta. 

Mathematical models of the human and animal placenta function have been proposed for at least~60 years~\citep[see discussions in][]{Battaglia1986,Aifantis1978,Gill2011,chernyavsky_2010}.
Previous models mainly focused either on a single villus scale or on the whole placenta scale \citep[different kinds of the flat wall exchanger model:][]{Bartels1962,Shapiro1967,Kirschbaum1969,Hill1973,Longo1972a,
Lardner1975,Groome1991,Wilbur1978,Gill2011}; 
several studies dealt with flow patterns~
\citep[co-orientation of maternal and fetal flows:][]{Battaglia1986,Bartels1962,Metcalfe1964,Shapiro1967,Faber1969,Kirschbaum1969,Guilbeau1970,Moll1972,
Schroder1982b}; 
other works represented the placenta as a porous medium~\citep{Erian1977,Schmid-Schonbein1988,chernyavsky_2010}, restricted, with one exception~\citep{chernyavsky_2010}, to one or two dimensions.
Some efforts have been devoted to understanding the relation between morphometric data and gas transfer in~1D in terms of diffusing capacity~(see~\citeauthor{Mayhew1986}, \citeyear{Mayhew1986} and references therein).

Better understanding of the transfer function of the placenta requires a model of the 3D geometry of the organ on a larger scale than a few villi, predictions of which could be compared 
to experimental data. %
We use a simplified engineering representation of the complex system in order to reveal the most relevant transport mechanisms.

The driving questions of our placenta modeling are:
\begin{itemize}

\item What are the relevant parameters~(geometrical and physiological) that govern oxygen transfer efficiency?

\item For a given set of physiological parameters (such as MBF velocity or oxygen content of blood), is there an \enquote{optimal} villous tree geometry which maximizes the oxygen uptake?

\end{itemize}

In the following, we present a simplified stream-tube placenta model~(STPM) of oxygen exchange. The mathematical equations governing oxygen transfer in the placentone are then numerically solved for different values of geometrical and physiological parameters. The results are compared to published histomorphometric measurements near term~\citep{Mayhew2000,Aherne1966,Nelson2009,Lee1995,Mayhew1993Conductance}.

\section{Mathematical model}

\subsection{Outline}

Our~3D~STPM is inspired by the observation that the human placenta resembles \enquote{a closed cubical room supported by cylindrical pillars running from floor to ceiling}~\citep{Lee1995}. 
The model consists of a large cylinder~(Fig.~\ref{figPlacentaModel3DRound}) representing a stream tube along which maternal blood flows~(Fig.~\ref{fig:PositioningOfTheGeoemtry}). This cylinder contains multiple smaller parallel cylinders which represent fetal villi~(terminal and mature intermediate), filled with fetal blood. In this paper, we consider these small cylinders to be of identical sizes, while their lateral spacings are randomly distributed~(with no cylinders overlap). This technical assumption can be easily relaxed in order to treat routinely stained 2D histological sections of the placenta.

\newcommand{\ThirdFigureCaption}{
\protectedSubref{figPlacentaModel3DRound}
The stream-tube model of oxygen exchange in a human placenta functional unit~(a placentone). Arrows show the flow of maternal blood. All the calculations are presented for a cylindrical stream-tube shape.
\protectedSubref{figPlacentaModel3DSquare}
An alternative~(square) geometry of a stream tube. 
As discussed in Appendix~\ref{appendix:sect:ValuesOfModelParameters}, the shape of the external boundary can be chosen arbitrary.
\protectedSubref{fig:PositioningOfTheGeoemtry}
A scheme of the placentone and location of the stream tubes in the placenta.
The dashed line schematically outlines the central cavity. Curved arrows on the right show maternal blood loosing oxygen while going from the central cavity to decidual veins. Curved lines on the left schematically show stream tubes of blood flow. The~STPM corresponds to one such tube unfolded; small straight arrows show the entrance point of the model.
The exchange is not modeled in the central cavity, but only after it, in the~MBF pathway. The concept that spiral arteries open into the~IVS near the central cavity corresponds to the current physiological views~\citep{chernyavsky_2010,benirschke_pathology}.
}
\begin{figure*}[tb]
\newcommand{\myFigHeight}{1.7in}
\centering \subfloat[]{
\includegraphics[
height=\myFigHeight, 
natheight=477, natwidth=912]
{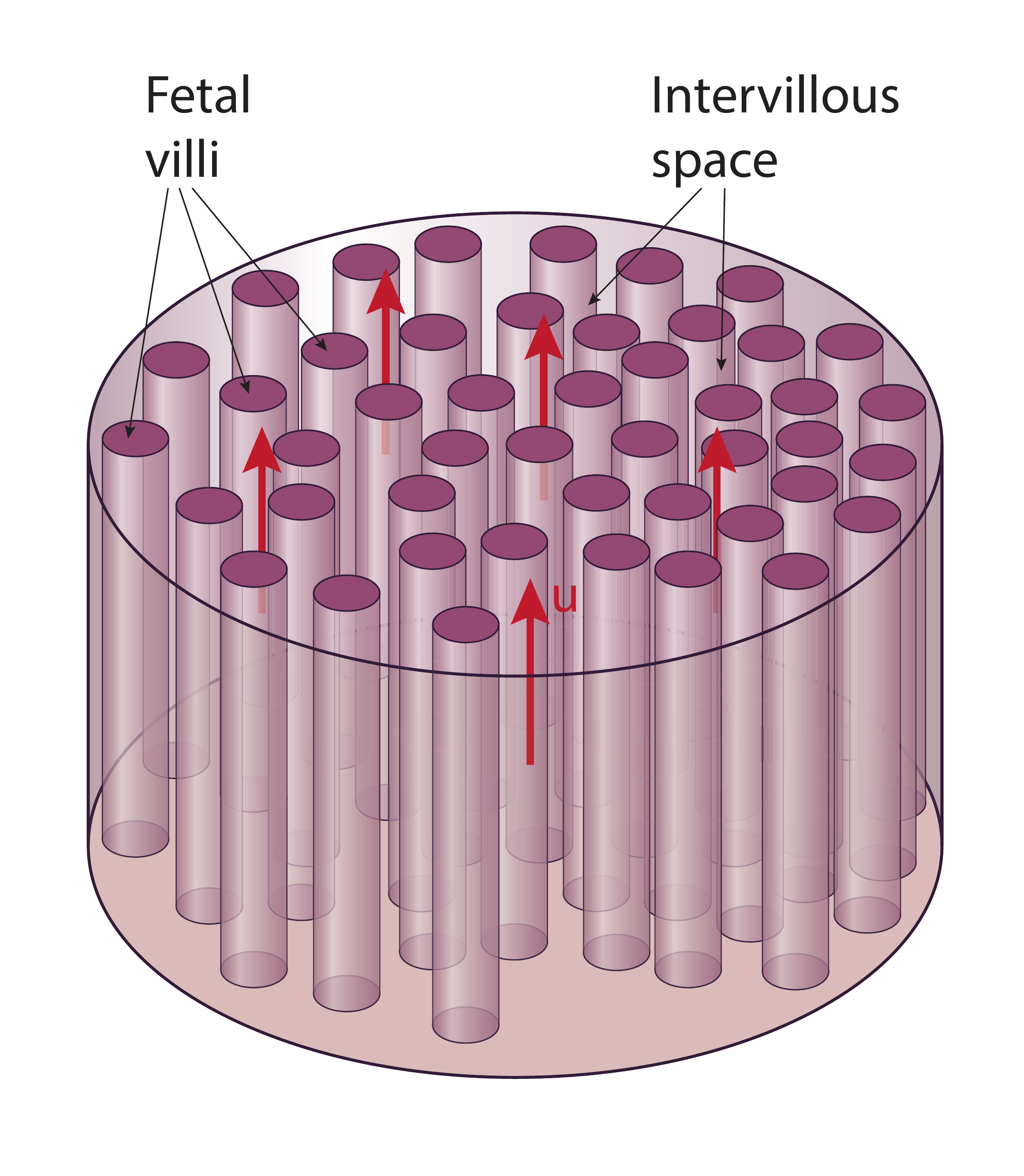}
\label{figPlacentaModel3DRound}
}
\subfloat[]{
\includegraphics[
height=\myFigHeight, 
natheight=477, natwidth=912]
{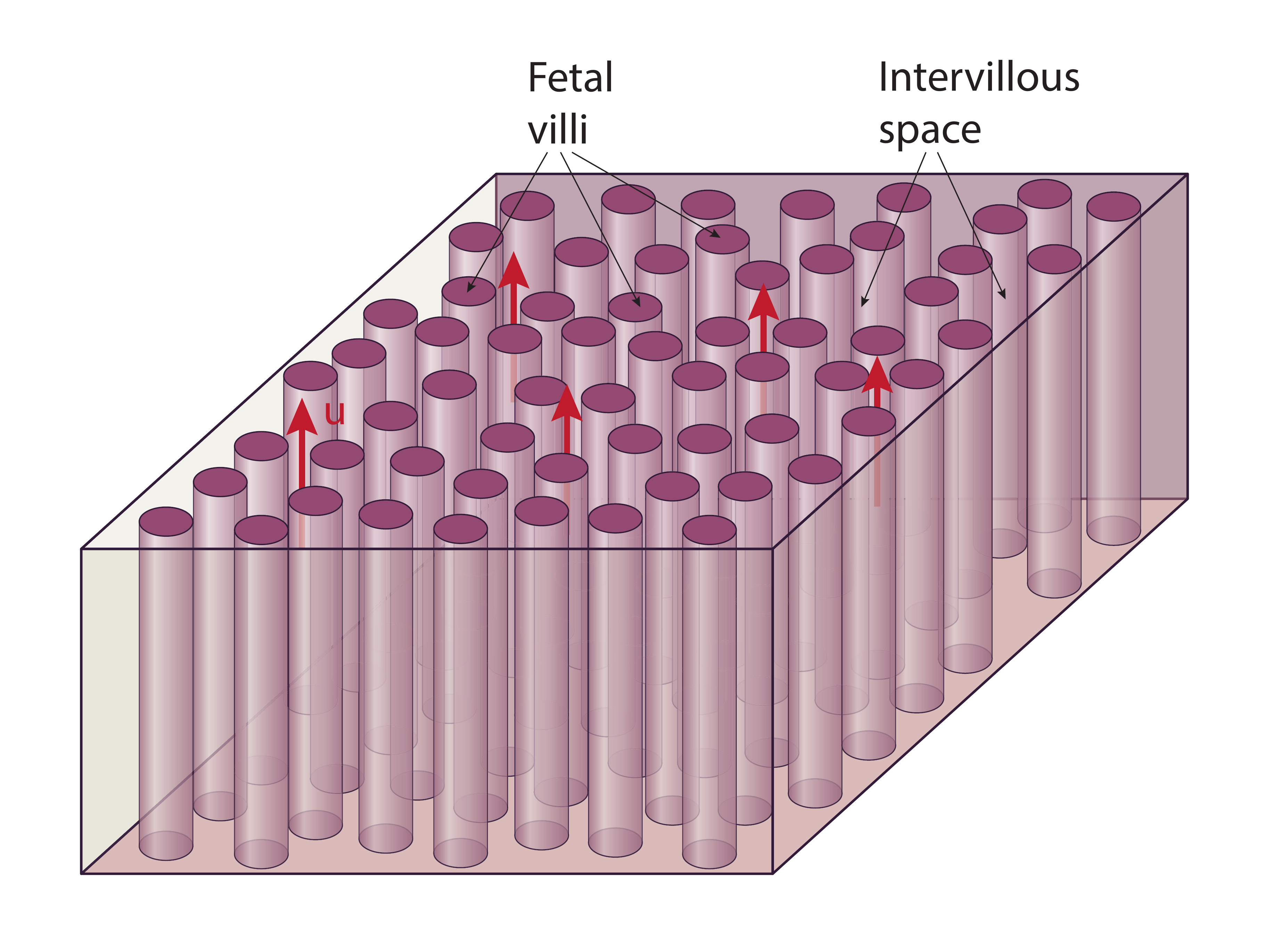}
\label{figPlacentaModel3DSquare}
}
\subfloat[]{
\includegraphics[
height=\myFigHeight,
natheight=1, natwidth=1]
{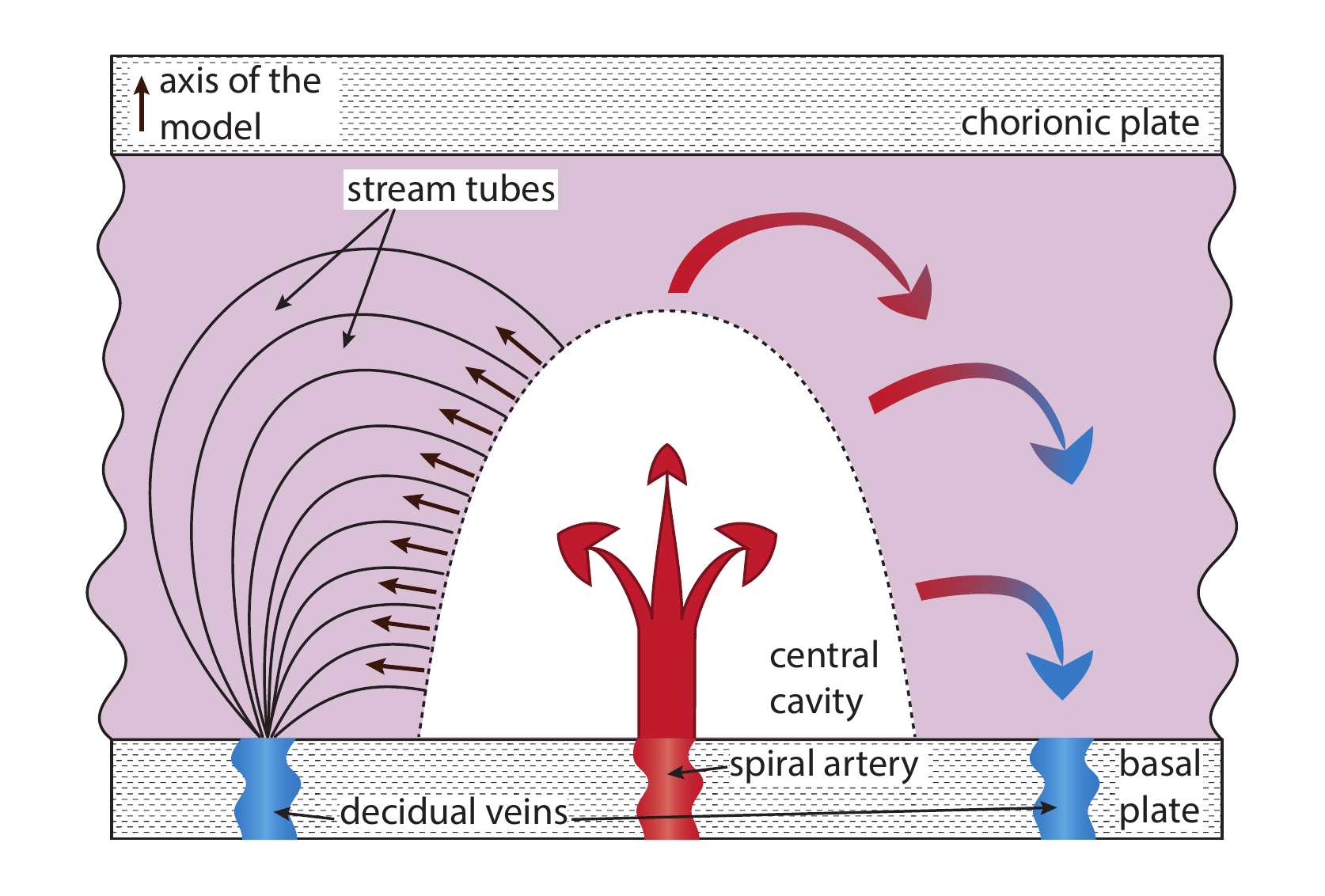}
\label{fig:PositioningOfTheGeoemtry}
}
\caption{
\ThirdFigureCaption
}
\label{fig:PlacentalModelBoth}
\end{figure*}

Our model relies on the following assumptions:
\begin{enumerate}
\item \emph{The large cylinder corresponds to an idealized unfolded stream tube of maternal blood}, extending from the boundary of the central cavity to a decidual vein~(Fig.~\ref{fig:PositioningOfTheGeoemtry}).
Every portion of arterial blood coming into placenta has its own stream tube, along which blood exchanges oxygen with fetal villi and gradually becomes venous. The contorted axis of a stream tube follows the maternal blood flow. The length of a tube can vary from one region to another, but is expected to be of the order of placental disk thickness~(see Appendix~\ref{appendix:sect:ValuesOfModelParameters} for more details);

\item \emph{The stream-tube has the same cross-section along its axis.}
In fact, we aim to base our STPM on histological slides~(like those shown in Fig.~\ref{fig:HealthyAndAbnormalPlacentaSections}), which give us only one section of a stream tube without any information about how this section changes along it. We then suppose that the same integral characteristics, such as the density or the perimeter of villi, are preserved along the same tube. %
This is obviously an oversimplification of the irregular~3D structure of the placenta, but it is the most straightforward assumption given the lack of the complete~3D geometrical data with all spatially resolved terminal and mature intermediate villi;

\item \emph{In any cross-section, oxygen is only redistributed by diffusion.} The gradient of oxygen dissolved in the blood plasma, not the gradient of oxygen bound to hemoglobin, determines the rate of oxygen diffusion in blood~\citep{Bartels1962}. 
As oxygen has low solubility in blood, linear solubility~\citep[Henry's law,][]{Prausnitz1998} is assumed;

\item \emph{Fetal blood is considered to act as a perfect oxygen sink}.  
This assumption means that we are studying oxygen partial pressure differences between the maternal and fetal blood. Additionally, it supposes that fetal circulation is efficient enough to rapidly carry away all oxygen that is transferred from maternal blood.
If villi are considered to be perfect sinks, there is no difference between concurrent, counter-current blood flow patterns or any other organization of the flow in this specific geometry. The features of the fetal blood flow can therefore be disregarded, except that fetal blood flow rate is assumed to be sufficient to maintain steep oxygen gradients for diffusion. In other words, the STPM separates the~IVS geometry from capillary geometry inside the villi. Resistance to oxygen transfer across the villus membrane, intravillous space and fetal capillary membranes %
are modeled by an effective feto-maternal interface of finite permeability calculated from the placental diffusing capacity;

\item \emph{Oxygen uptake occurs at the feto-maternal interface, i.e. at the boundaries of the small cylinders in Fig.~\ref{figPlacentaModel3DRound}. It is directly proportional to the interface permeability and to the oxygen concentration on the maternal side of the interface}. This statement is a reformulation of Fick's law of diffusion, in which the rate of oxygen transfer across the interface is proportional to the difference of partial pressures ($\pONew$) on both sides of the interface;

\item \emph{Maternal blood flow is considered to be laminar with no liquid-walls friction~(slip boundary conditions at all surfaces), so that the velocity profile in any cross-section is flat}. This assumption is supported by calculations obtained for capillary-tissue cylinders in brain~\citep{Reneau1967}. The brain model has been used to compare the effect of non-slip boundary conditions~(and thus a non-flat velocity profile) versus slip boundary conditions~(and thus a flat velocity profile) on the distribution of~$\pONew$ in a Krogh's type cylindrical tissue layer. The difference between the two cases in~$\pONew$ distribution was less than~$\unit[10]{\%}$. This strong assumption is discussed later in the text;

\item \emph{Oxygen-hemoglobin dissociation curve is linearized} in order to simplify the resolution of equations;

\item \emph{Erythrocytes are uniformly distributed in the~IVS;}

\item \emph{The solution is stationary}, i.e. placental oxygen flow remains constant over time;

\item \emph{Blood flow matching does not occur}, i.e. there is no influence~(feedback) of oxygen uptake on the parameters of the model~
\citep[such as~MBF velocity, see][]{Talbert2004}. No redundancy of some of the fetal cylinders when maternal~IVS blood flow is reduced is taken into account.

\end{enumerate}

\subsection{Solution of the model}

\subsubsection{Characteristic time scales of the transfer processes in the placenta}
\label{sect:TimeScales}

We identify three different physical transfer processes in the placenta, each of which operates on a characteristic time scale: hydrodynamic blood flow through the~IVS characterized by an average velocity~$u$ and transit time~$\tauPas$; diffusion of oxygen with characteristic time~$\tauD$; and equilibration between oxygen bound to hemoglobin and oxygen dissolved in the blood plasma with characteristic time~$\tauP$. This last thermodynamic equilibrium is described as equal partial pressure of oxygen in both states.
The three times can be estimated as follows:
\begin{itemize}
\item $\tauPas$, the transit time of blood through the~IVS, is of the order of~\unit[27]{s} from the results of angiographic studies at term~\citep[see Appendix~\ref{appendix:sect:ValuesOfModelParameters}]{Burchell1967};

\item $\tauD$, reflecting oxygen diffusion over a length~$\delta$ in the~IVS is~
$\tauD\sim\delta^2/D$, where~$D=\unit[(1.7\pm0.5)\cdot10^{-9}]{m^2/s}$ is oxygen diffusivity in the blood plasma~\citep{Moore2000a}.
Either from calculations~\citep{Mayhew2000}, or directly from %
normal placenta sections~(Fig.~\ref{fig:HealthyAndAbnormalPlacentaSections}), mean width of an~IVS pore can be estimated as~$\delta\sim\unit[80]{\mu m}$, yielding~$\tauD\sim\unit[4]{s}$;

\item $\tauP$, an equilibration time scale, which includes the characteristic diffusion time for oxygen to reach Hb~molecules inside a red blood cell~\citep[{$\sim\unit[10]{ms}$}][]{Foucquier2013} and a typical time of oxygen-hemoglobin dissociation~($\sim\unit[20]{ms}$ for the slowest process, see~\citealp{Yamaguchi1985}). Together these times sum to~$\tauP\sim\unit[30]{ms}$.
\end{itemize}
These three characteristic times are related as follows:~$\tauP\ll\tauD\lesssim\tauPas$. This relation suggests that oxygen-hemoglobin dissociation can be considered instantaneous as compared to diffusion and convection; the latter two, by contrast, should be treated simultaneously. Further details of the solution of the model can be found in Appendix~\ref{appendix:sect:MathematicalFormualtion}.

\subsubsection{Parameters of the model}
\label{sectModelParameters}

\begin{table}[tb]
\caption{\rm
	Parameters of the human placenta used in the calculations. Details of the parameters calculation can be found in Appendix~\ref{appendix:sect:ValuesOfModelParameters}. No experimental error estimations were available for~$u$ and $L$.}
	\smallskip
	\label{tblCalculationsParameters}
	    \centering	
		\begin{tabularx}{\columnwidth}{@{}p{2.0in}lr@{\ }l@{}}
			\toprule
			Parameter & Symbol & Mean & $\pm$ SD
			\\
			\midrule
			Maximal Hb-bound oxygen concentration at~\unit[100]{\%} Hb saturation,~$\unit{mol/m^{3}}$	&	$\cMax$	&	7.30 & $\pm$ 0.11
			\\
			Oxygen-hemoglobin dissociation constant	&	$B$	&	94 & $\pm$ 2
			\\
			Concentration of oxygen dissolved in blood at the entrance to the~IVS,~$\unit[10^{-2}]{mol/m^{3}}$	&	$c_0$	&	6.7 & $\pm$ 0.2
			\\			
			Oxygen diffusivity in blood,~$\unit[10^{-9}]{m^2/s}$	&	$D$	&	1.7 & $\pm$ 0.5
			\\			
			Effective villi radius,~$\unit[10^{-6}]{m}$	&	$r$	&	41 & $\pm$ 3
			\\
			Permeability of the effective materno-fetal interface,~$\unit[10^{-4}]{m/s}$	&	$w$	&	2.8 & $\pm$ 1.1
			\\			
			Placentone radius,~$\unit[10^{-2}]{m}$	&	$R$	&	1.6 & $\pm$ 0.4
			\\
			Radius of stream tube used in numerical calculations,~$\unit[10^{-4}]{m}$	&	$\RNum$	&	6 &
			\\		
			Velocity of the maternal blood flow,~$\unit[10^{-4}]{m/s}$	&	$u$	&	6 &
			\\
			Stream tube length,~$\unit[10^{-2}]{m}$	&	$L$	&	1.6 &
			\\
			\bottomrule
		\end{tabularx}
\end{table}

The parameters necessary for the construction of the model are listed in Table~\ref{tblCalculationsParameters}, while their calculation is reported in Appendix~\ref{appendix:sect:ValuesOfModelParameters}. 
Errors in the experimental parameters were estimated in cases when such information was available from the source; no error assumptions were made in other cases.
Relatively small stream-tube radius~$\RNum$ has been used to speed up the computations, but all results have been rescaled to the placentone radius~$R$ by multiplying the uptake by~$R^2/\RNumSquared$.

Note that besides the parameters traditionally used for modeling 1D exchange in the human placenta, namely: incoming~($c_0$) and maximal oxygen concentrations~($\cMax$), diffusivity of oxygen in blood~($D$), solubility of oxygen in blood~($\kHenry$), oxygen-hemoglobin dissociation coefficient~($\bSixty$), density of blood~($\rhoBl$), length of the blood flow path~($L$), a characteristic size of the exchange region~($R$) and~the local permeability of the feto-maternal interface~($w$) or the integral diffusing capacity of the organ~($\DP$), the construction of the STPM required several additional parameters to account for a~2D placental section structure~\citep[compare, for example, with][]{Power1972a}. These are: (i)~the linear maternal blood flow velocity in the exchange region~($u$, in \unit{m/s}), since the usual integral flow (in \unit{ml/min}) is not enough; (ii)~the volumetric density~($\phi$) of small fetal villi in a given region; and (iii)~the effective radius of fetal villi~($r$), which accounts for the non-circular shape of fetal villi by providing the right volume-to-surface ratio.

\subsubsection{Numerical simulation}

\newcommand{\FourthFigureCaption}{
A range of villi densities for which oxygen uptake has been calculated. The number of villi and the corresponding villi density is displayed above each case. Simulations have been carried out for all geometries in the range~$N=1\text{--}160$.
}
\begin{figure}[tb]
\center{\includegraphics[
width=3.5in, 
natheight=477, natwidth=912]
{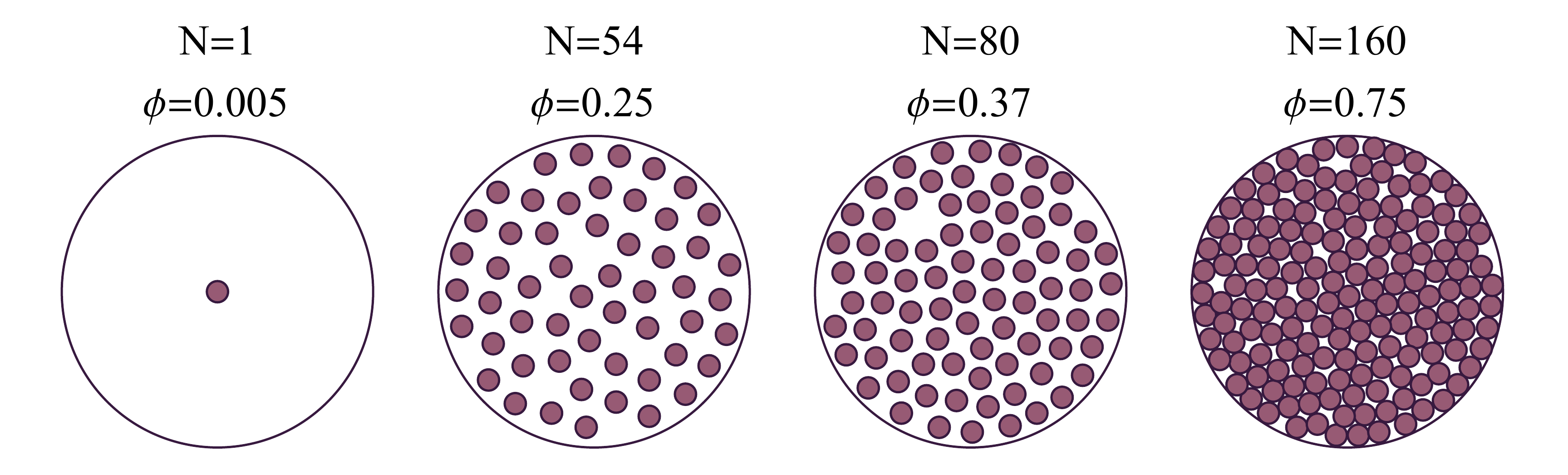}}
\caption{
\FourthFigureCaption
}
\label{fig:CalculatedGeometries}
\end{figure}

The diffusion-convection equation~\fm{fmGeneralEquationLinear} %
is solved by a finite elements method in Matlab\registeredTrademark\ for various 2D villi distributions~(Fig.~\ref{fig:CalculatedGeometries}). 
Villi distributions are generated in the range of villi densities~$\phi$ from~0.005 to~0.75.
Villi density is defined as the ratio of small villi cross-sectional area~($\SSmallVilli$) to the total area of the stream tube cross-section~($\sTot$): $\phi=\SSmallVilli/\sTot$. These densities correspond to a range of villi numbers~$N$ from~1 to~160 for a stream-tube of the radius~$\RNum$. 
The upper boundary of the range has been chosen close to the maximal packing density of circles that one can achieve without resorting to special packing algorithms~\citep[][]{Specht2009}.


\section{Results}

Figure~\ref{fig:UptakeLVilliDensity} shows the dependence of oxygen uptake of a single placentone on the length of the stream tube. One can see that the villi density which provides the maximal oxygen uptake, depends on the stream tube length, all other parameters remaining fixed.
In other words, \emph{the most efficient geometry} (the one which gives the highest uptake for a fixed length,~MBF velocity and incoming oxygen concentration) \emph{is different for different stream tube lengths}. The villi density corresponding to the maximal uptake will be called \enquote{optimal}. 
The fact that there is an optimal villi density can be clearly seen in Fig.~\ref{fig:UptakeVilliDensityFixedL}, which shows the variation of oxygen uptake with villi density for a fixed stream tube length. 

\newcommand{\FifthFigureCaption}{
Oxygen uptake of a single placentone as a function of several geometrical parameters.
\protectedSubref{fig:UptakeLVilliDensity}
Oxygen uptake as a function of the stream-tube length~$L$ for fixed villi densities~$\phi$.
Various symbols represent the villi densities of Fig.~\ref{fig:CalculatedGeometries}.
\protectedSubref{fig:UptakeVilliDensityFixedL}
Oxygen uptake as a function of villi density~$\phi$ for three lengths~$z$: $L/3$, $2L/3$, and $L$. The peak uptake moves to smaller villi densities when the length~$z$ increases. %
Oxygen uptake has been calculated for~$\RNum$ and rescaled to placentone radius~$R$ by multiplying by~$R^2/\RNumSquared$. Transit time of maternal blood through the placentone~($\tauPas=z/u$), blood inflow per placentone~($u\pi R^2 (1-\phi)$) and oxygen uptake in~$\unit{cm^3/min}$ are shown on the additional axes to simplify comparison of the results with physiological data.
}
\begin{figure*}[tb]
\newcommand{\picHeight}{0.2}  
\vspace{-4ex} \centering 
\subfloat[]{
\includegraphics[height=\picHeight\paperwidth, natheight=477, natwidth=912]
{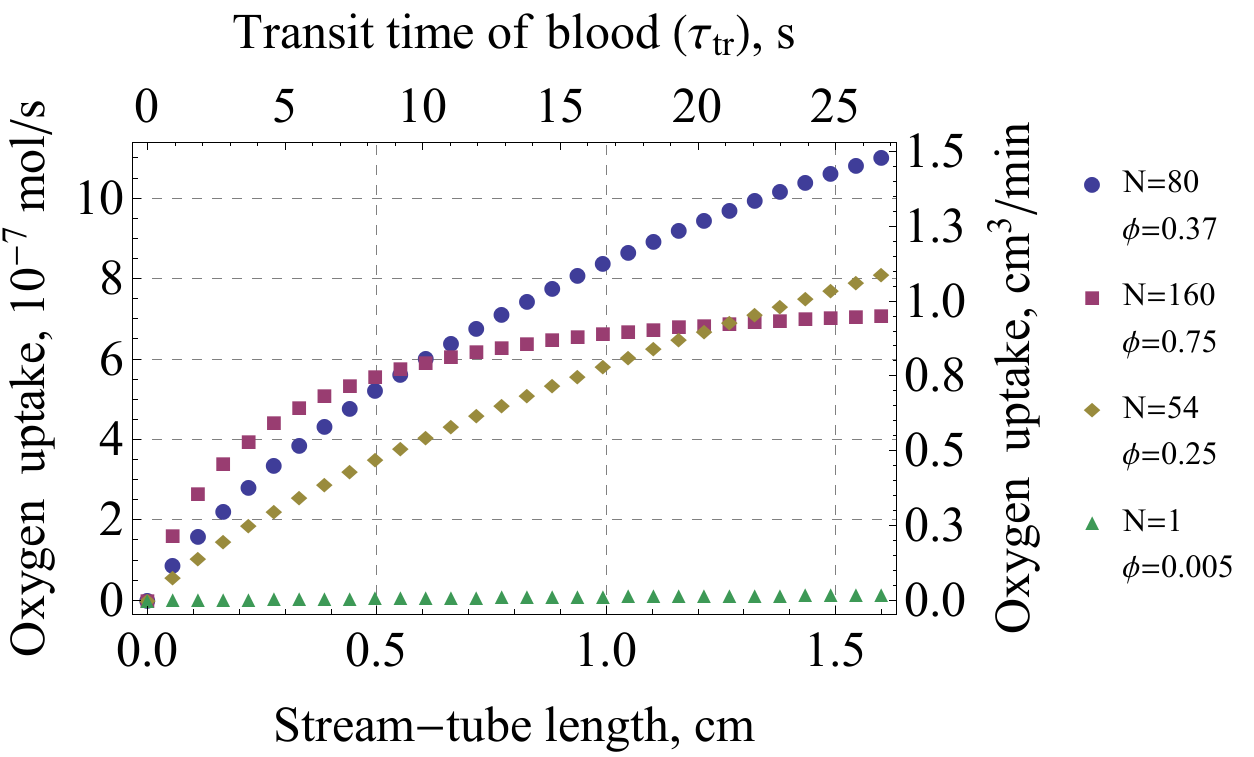}
\label{fig:UptakeLVilliDensity} }
\hspace{1ex}
\subfloat[]{
\includegraphics[height=\picHeight\paperwidth, natheight=477, natwidth=912]
{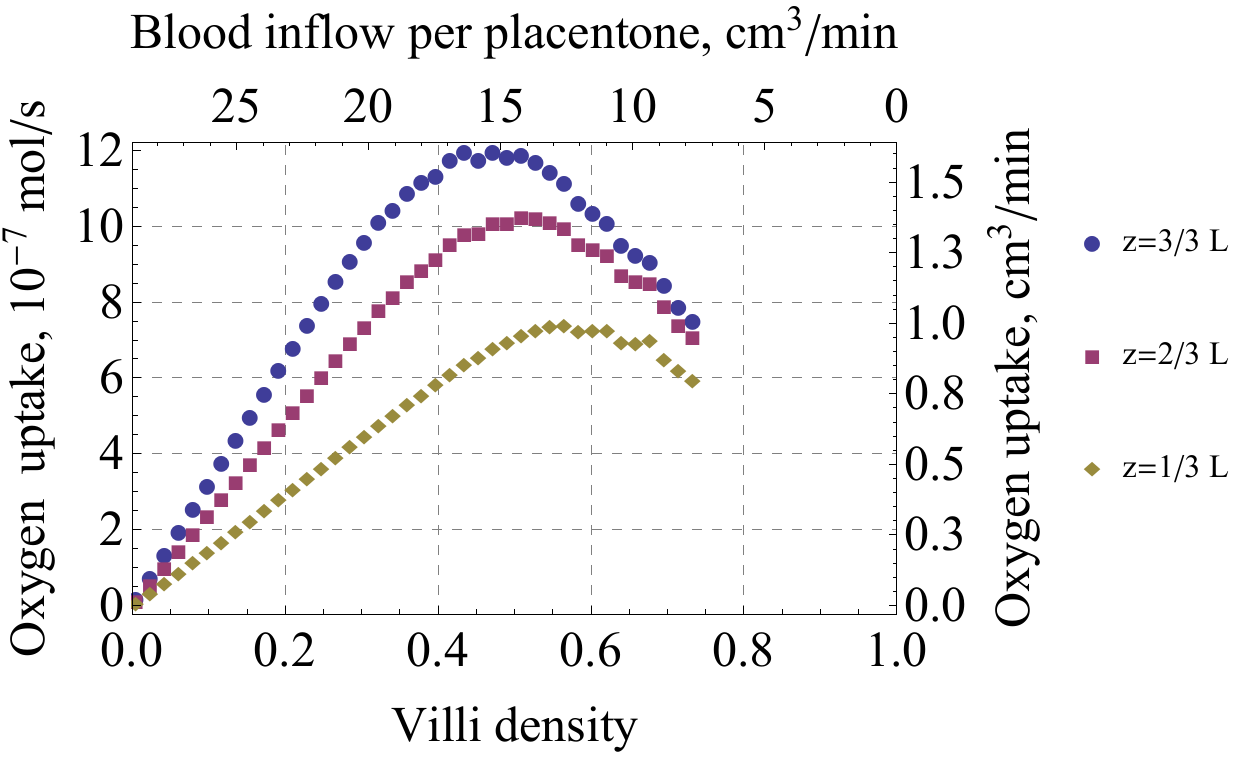}
\label{fig:UptakeVilliDensityFixedL} }

\caption{
\FifthFigureCaption
}
\label{fig:UptakeResults}
\end{figure*}

Both the maximal oxygen uptake and the corresponding optimal villi density depend on the average~MBF velocity and on the stream-tube length~$L$, which are the parameters the most susceptible to variations in different placentas and for which no error estimations were available (see Appendix~\ref{appendix:sect:ValuesOfModelParameters}). 
If empirical distribution of~MBF velocities in the~IVS of the human placenta were known, with the help of Fig.~\ref{fig:VilliDensityVelocity}, \emph{one could estimate the optimal villi density that would result in the maximal uptake} in any given region. Conversely, if one knows the average villi density in a placenta region (e.g., from histological slides)  and if optimal placenta function is assumed, \emph{one can estimate the~MBF velocity} in that region. The corresponding maximal oxygen uptake can be determined from Fig.~\ref{fig:MaximalUptakeVelocity}. The dependence of the optimal characteristics on the stream-tube length is shown in Figs~\ref{fig:VilliDensitySystemLength}, \ref{fig:MaximalUptakeSystemLength}.

\newcommand{\SixthFigureCaption}{
\protectedSubref{fig:VilliDensityVelocity}, \protectedSubref{fig:MaximalUptakeVelocity}: %
Dependence of the optimal villi density~\protectedSubref{fig:VilliDensityVelocity} and the maximal oxygen uptake~\protectedSubref{fig:MaximalUptakeVelocity}
on the~MBF velocity at a fixed length~$L$ for a single placentone. %
\protectedSubref{fig:VilliDensitySystemLength}, \protectedSubref{fig:MaximalUptakeSystemLength}:~%
Dependence of the optimal villi density~\protectedSubref{fig:VilliDensitySystemLength} and the maximal oxygen uptake~\protectedSubref{fig:MaximalUptakeSystemLength}
on the stream-tube length~$L$ %
for a single placentone. %
Dashed lines show the expected average stream tube length~$L=\unit[1.6]{cm}$ and~MBF velocity~$u=\unit[6\cdot10^{-4}]{m/s}$. Transit time of maternal blood through the placentone~($\tauPas=L/u$), blood inflow per placentone~($u\pi R^2 (1-\phi)$) and oxygen uptake in~$\unit{cm^3/min}$ are shown on additional axes to simplify comparison of the results with physiological data.
}
\begin{figure*}[tb]
\newcommand{\picWidth}{0.37}
\vspace{-4ex} \centering 
\subfloat[]{
\hspace{-2.4ex}
\includegraphics[width=\picWidth\paperwidth, natheight=1804, natwidth=1693]
{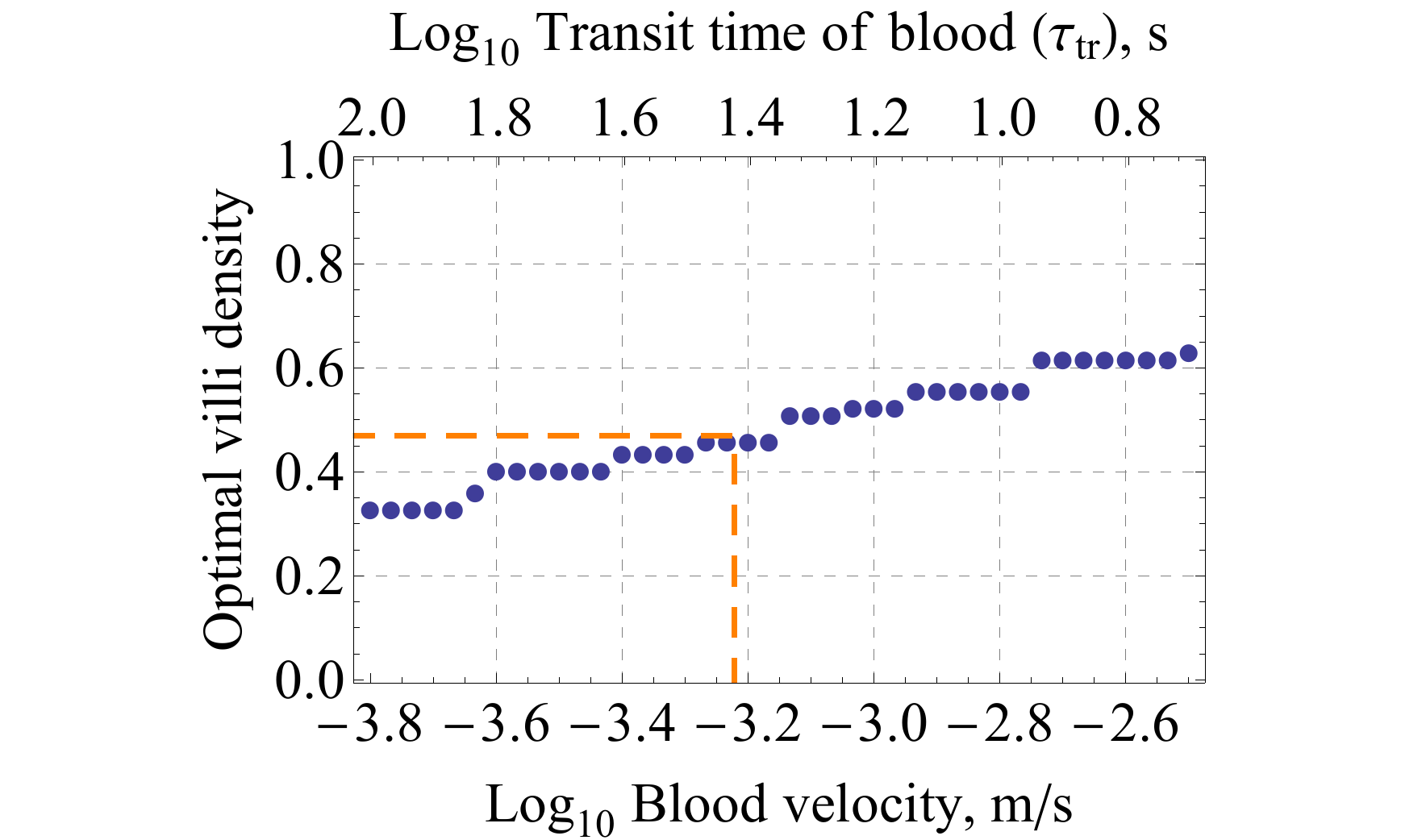}
\label{fig:VilliDensityVelocity} }
\subfloat[]{
\includegraphics[width=\picWidth\paperwidth, natheight=1804, natwidth=1693]
{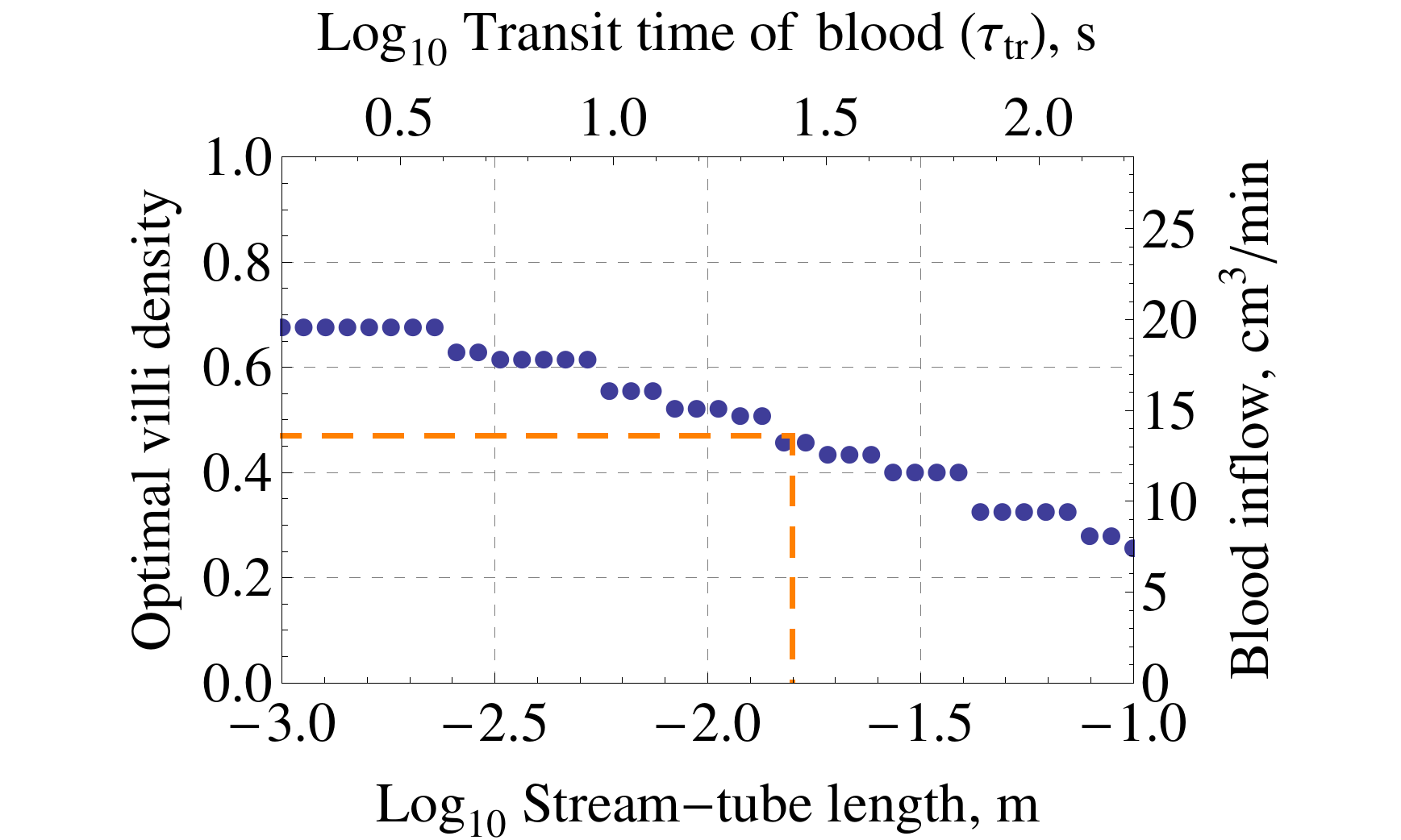}
\label{fig:VilliDensitySystemLength} }
\vspace{-2ex}
\subfloat[]{
\includegraphics[width=\picWidth\paperwidth, natheight=1804, natwidth=1693]
{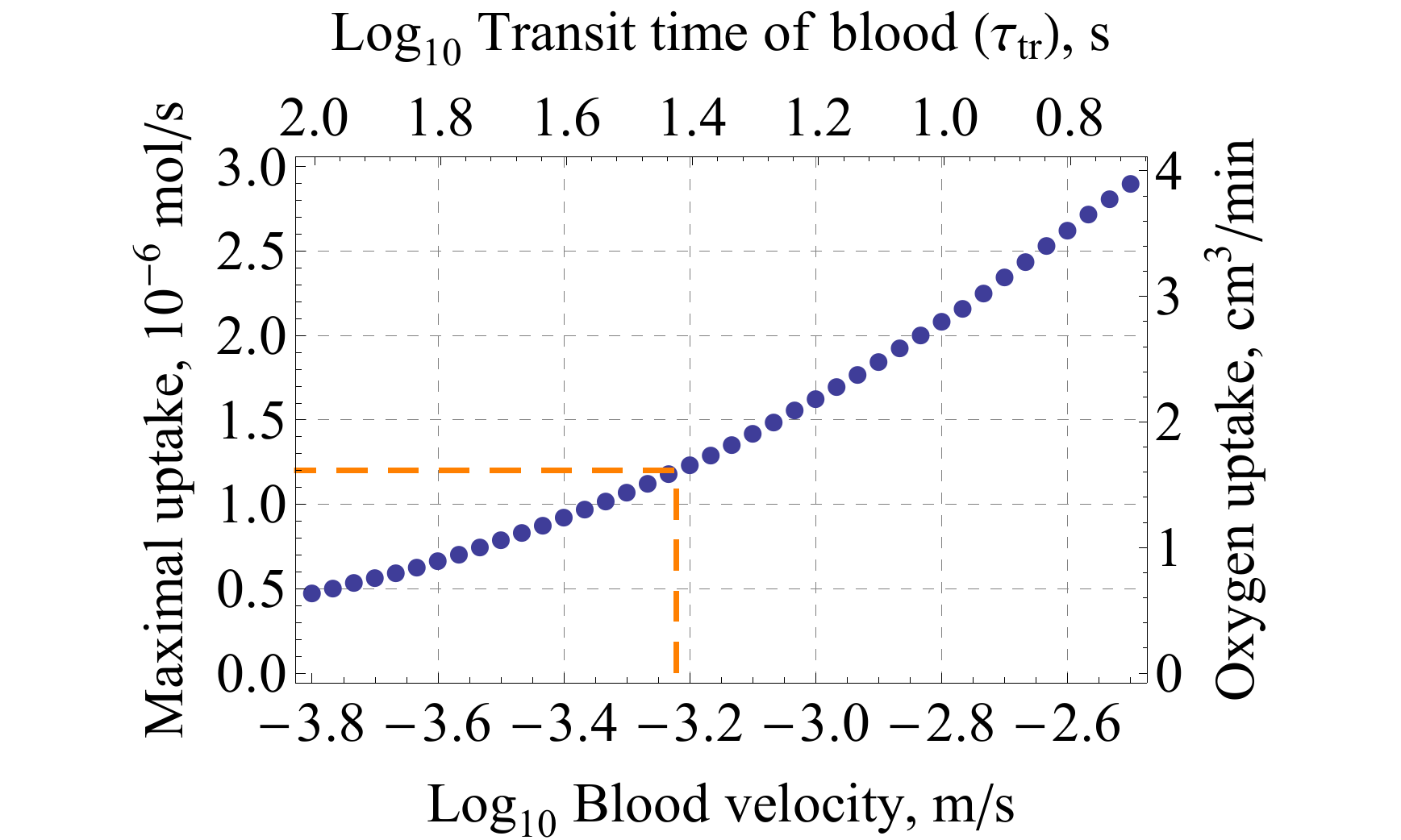}
\label{fig:MaximalUptakeVelocity} }
\subfloat[]{
\hspace{-3.2ex}
\includegraphics[width=\picWidth\paperwidth, natheight=1804, natwidth=1693]
{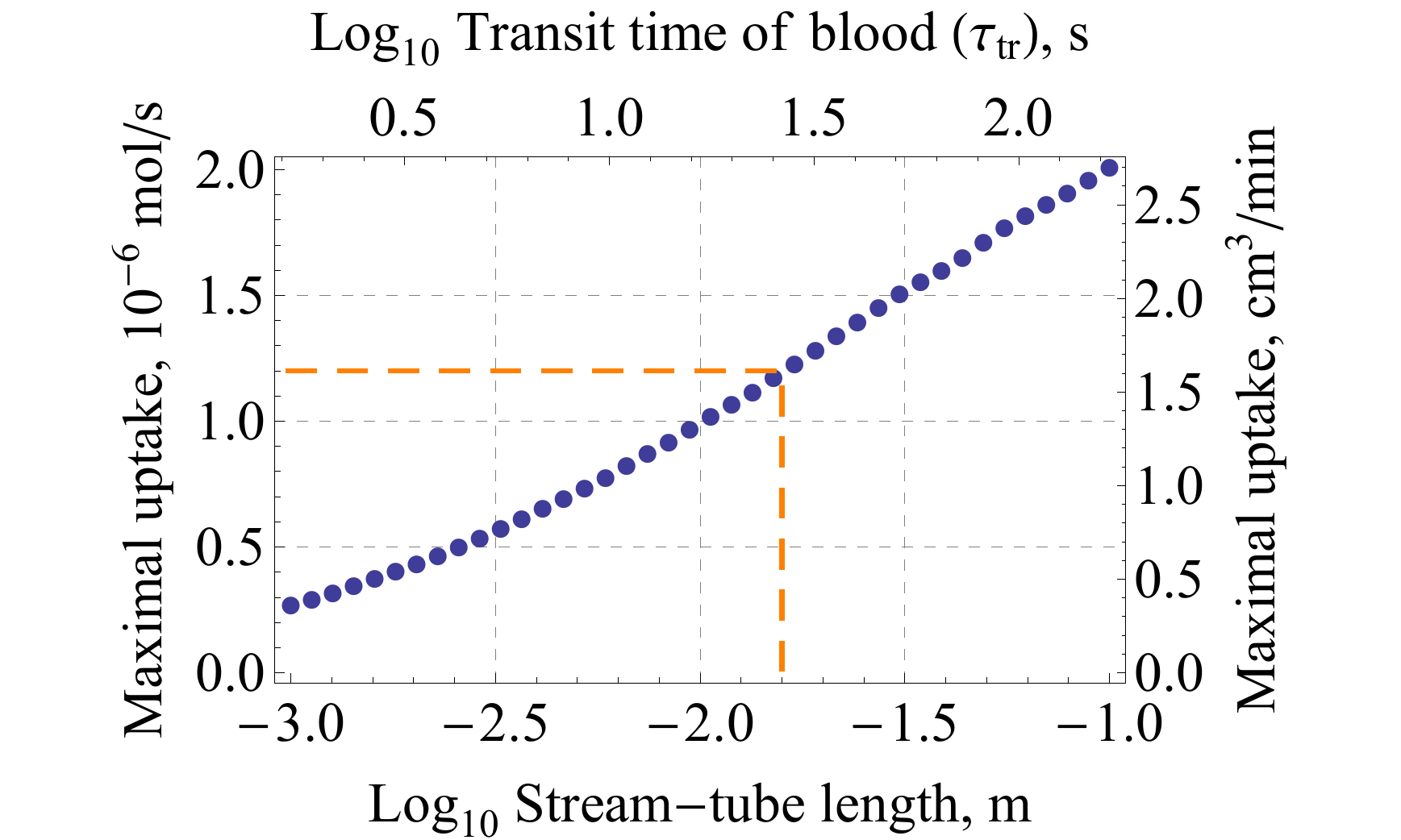}
\label{fig:MaximalUptakeSystemLength} }

\caption{
\SixthFigureCaption
}
\label{fig:ResultsDependenceOnVelocity}
\end{figure*}

Our calculations show that for a healthy placenta~(with the parameters given in Table~\ref{tblCalculationsParameters}), the optimal villi density is~$\phi_0=0.47\pm0.06$ and the corresponding maximal oxygen uptake is~$\FMax=\unit[(1.2\pm0.6)\cdot10^{-6}]{mol/s}\approx\unit[1.6\pm0.8]{cm^3/min}$. Details of the provided error estimations can be found in Appendix~\ref{appendix:error_estimation}. It is worth nothing that to the precision of the calculations, no dependence of the results on the diffusivity~$D$ of oxygen was observed. Note also that since no experimental error data was available for~$L$ and~$u$, these variables were not included in error calculation. The confidence intervals for~$\phi_0$ and~$\FMax$ are hence likely to be larger than the provided values. 

\section{Discussion}

We have developed the~STPM of oxygen exchange in the human placenta that relates oxygen uptake to cross-section histomorphometry obtained from routinely prepared placenta slides. 
Our goal was %
to provide a model capable of identifying cases of suboptimal oxygen transfer from maternal to fetal blood by histological analysis of a placenta post-partum. %
It was shown that there is always an optimal villi density, at which oxygen uptake is maximal, all other parameters being fixed.

The optimal density is determined by a competition of two factors. On one hand,
larger villi density leads to larger exchange surface and thus larger uptake.
On the other hand, larger villi density implies that less blood is coming into the placentone per unit of time, hence a lower uptake rate.
For a given set of the placenta-specific parameters~(Table~\ref{tblCalculationsParameters}), our model allows one to calculate the optimal villi density for a given placenta or placenta region.


\subsection{Comparison to experimental data}


Our calculated optimal density can be compared to experimental data. The villi density~$\phi$ can be determined from published histomorphometrical data according to the formula~$\phi= \SSmallVilli/ (\SSmallVilli+\SM)$,
where~$\SSmallVilli$ is the cross-sectional area of terminal and mature intermediate~(small) villi, and~$\SM$ is the cross-sectional area of the~IVS. 
In the literature,~$\SSmallVilli$ and~$\SM$ are traditionally obtained as surface area fractions of a~2D section and then multiplied by the volume of the placenta to yield volumes of small villi~($\vSmallVilli$) and~IVS~($\vIVS$). Villi density can be calculated by the same formula from these volumes:~$\phi=\vSmallVilli/(\vSmallVilli+\vIVS)$~
(Table~\ref{tbl:HistomorphometricPhiValues}).
In cases when the fraction of terminal and mature intermediate villi in the total villi volume was not available, we have used the average fraction of small villi volume of~$\alpha\approx0.665$ in the total volume of the villi~\citep{Sen1979}. Note that these experimental data refer to placentas at term since a morphometrical study of a placenta can be only performed post-partum.

\begin{table}[tbh]
\newcommand{\tblWidth}{\columnwidth}
\caption{
\rm
Villi density~($\phi$) as calculated from histomorphometrical data for small villi~($\vSmallVilli$) and IVS~($\vIVS$) volumes of placentas at term obtained from normal pregnancies.}
	\smallskip
	\label{tbl:HistomorphometricPhiValues}
	    \centering
		\begin{tabularx}{\tblWidth}{@{}XXXl@{}}
		\toprule
		$\vSmallVilli$, $\unit{cm^3}$ & $\vIVS$, $\unit{cm^3}$ & $\phi$	&	Source
		\\
		\midrule
		130 & 170 & 0.43 & \citet{Mayhew2000}
		\\
		220 & 220 & 0.50 & \citet{Nelson2009}
		\\
		190 & 170 & 0.53 & \citet{Aherne1966}
		\\
		110 & 180 & 0.38 & \citet{Lee1995}
		\\
		150 & 170 & 0.47 & \citet{Mayhew1993Conductance}
		\\
		\bottomrule
		\end{tabularx}\par
\end{table}

The optimal villi density~$0.47\pm0.06$ obtained in our model is consistent with the experimental data which yield~$\phi=0.46\pm0.06$~(mean~$\pm$ SD). %
%
The quantitative agreement of the model's predictions with physiological densities allows us to speculate that normal placentas are optimal oxygen exchangers, while pathological placentas~(e.g. diabetic or pre-eclamptic) behave sub-optimally. Although such an argument is plausible, further analysis is required.
Evidence from sheep specifically demonstrates a large reserve capacity for fetal oxygen supply, suggesting that variation in villi density and~MBF velocity might be a normal tolerated biological process~\citep{Carter1989}. A similar behavior has been demonstrated earlier in human lungs, where lung geometry is not optimal under normal conditions, but its reserve capacity is used during exercises~\citep{Sapoval2002}. 

The experimental morphometrical analysis of the villi density in placenta sections post-partum may not represent precisely the \emph{in vivo} situation due to the cease of maternal and fetal blood flows and to the fixation procedure.
Because of these effects, the histological villi density seen post-partum may overestimate the \emph{in vivo} villi density.
However, for the moment, there is no generally acknowledged method which would allow to correct for these changes.

\subsection{Comparison to the porous medium model}

Our optimal villi density can be compared to predictions of the porous medium mo\-del~\citep[PMM,][]{chernyavsky_2010}.
\citeauthor{chernyavsky_2010} have obtained a value of the optimal fraction of \enquote{villous material}~$\phiC$ close to~$0.3$. This value has to be translated into~$\phi$ obtained in the~STPM to allow comparison.
Defining~$\phiC$ as a fraction of villi volume in the total volume of~IVS and villi, and~$\phi$ as a fraction of \emph{small} villi volume in the total volume encompassing IVS and \emph{small} villi:
\begin{linenomath}
\begin{align*}
&\phiC=\vVil/(\vVil+\vIVS),
 &&\phi=\vSmallVilli/(\vSmallVilli+\vIVS),
\end{align*}
\end{linenomath}
one obtains the following translation formula:~$\phi=\alpha/(\alpha-1+1/\phiC)\approx0.22$, where again~$\alpha\equiv\vSmallVilli/\vVil\approx0.665$ is the average fraction of small villi in the total villi volume~\citep{Sen1979}. Comparison of this value to the average experimental value~$\phi=0.46\pm0.06$ demonstrates a twofold underestimation of the observed villi density in the~PMM. The authors themselves admit that this value corresponds rather to high-altitude or pathological pre-eclamptic placentas than to healthy ones. A possible reason of this mismatch is that in the~PMM the optimal villi density appears as a trade-off between blood flow resistance and uniform uptake capacity of an unspecified passively transported substance. We propose a different explanation presenting \emph{the optimal villi density as a trade-off between the absorbing surface of fetal villi and the incoming flow of oxygen}. In this claim, we underline the role of the absorbing surface~(which is hard to define in the~PMM), but also suggest the importance of oxygen transport as compared to the passive transport of an unspecified substance. 

The two models offer two different points of view on modeling the human placenta, which we discuss below.
\begin{enumerate}
\item
Representing villi as a uniform and isotropic porous medium in the~PMM ignores the placenta geometry.
The only geometrical parameter left~(volume fraction of the \enquote{villous material}, analogous to villi density) masks independent roles played by the absorbing surface as well as by villi shapes and sizes. With no uptake surface, the~PMM requires assuming a uniform uptake proportional to the~IVS volume fraction. At the same time, experimental results show a considerable difference in the absorbing surface between healthy and pathological cases~\citep{Mayhew2000}. 

Although the full~3D structure of the human placenta at the microscopic scale of individual villi is not available from the experiment, its~2D sections can be obtained~(Fig.~\ref{fig:HealthyAndAbnormalPlacentaSections}). Both the~PMM and~STPM can rely on these slides in terms of villi density, but the present model also accounts for the absorbing surface by using an effective villi radius~(see Appendix~\ref{appendix:sect:ValuesOfModelParameters}) and can take a whole 2D section as input data~(the last option requires development of image analysis techniques). In the~PMM, the analysis cannot go beyond the villi density;

\item
Simple uniform uptake kinetics~(ignoring interaction with hemoglobin) used in the~PMM can only account for passive transport of certain metabolites, but it is unable to describe exchange of respiratory gases as it ignores~Hb-binding~(for instance, about \unit[99]{\%} of oxygen comes into the placenta in the bound form). At the same time, the~STPM can be used for oxygen and carbon dioxide transport as well as for passive substances transport.

\item
The use of slip boundary conditions yielding a flat velocity profile in the~STPM needs further discussion. Ideally, a full model of the placenta transport function would first require the full high-resolution~3D placenta structure to compute the velocity field by solving the Navier-Stokes equation with non-slip boundary conditions at the villous tree surface. After that, oxygen uptake could be calculated using convection-diffusion equations with the obtained velocity profile. Since modern experimental techniques can neither acquire the placenta structure with a sufficient spatial resolution for such calculations, nor provide direct experimental measurements of the blood velocity field, one must resort to simplifications. One way is to ignore the geometrical structure of the organ and substitute it by a uniform isotropic porous medium as in the~PMM. We chose an alternative way using geometrical information available from~2D histological slides.

From the hydrodynamic point of view, the non-slip boundary conditions would be the most natural for modeling blood flow in a real high-resolution 3D placenta structure.
Since the blood flow path in a real placenta geometry is expected to be irregular, \emph{no developed velocity profile} is expected along the flow, and different sections of the placenta may present different velocity profiles. These real velocity profiles, however, can be far from that of a \emph{developed} flow with non-slip boundary conditions. 
As a consequence, the advantage of choosing the non-slip boundary conditions over the slip ones in our already simplified structure is not so clear.
In other words, although slip and non-slip boundary conditions may lead to different quantitative predictions, it is difficult to say which prediction will be closer to the results in a real~3D placenta geometry. This statement can be verified either by modeling the blood flow in a full placenta geometry~(unavailable by now), or by confronting models predictions to physiological observations. From the latter point of view, our model with slip boundary conditions predicts the optimal villi density which is comparable to the experimentally observed one. A comparison with predictions from the same simplified structure with non-slip boundary conditions presents an interesting perspective. Note that the blood velocity used in our model should be understood as an \emph{average} velocity across the stream tube or a placenta region.



On the contrary, PMM implicitly takes into account the non-slip boundary conditions~(friction at the boundary) by using Darcy's law. However, the~PMM has similar limitations as it obtains a developed blood flow which may be far from the real non-developed one. Moreover, qualitative distribution of the flow in the placentone obtained in the~PMM remains approximate because: (i) Darcy's law has never been proved valid for the human placenta due to the lack of experimental data; (ii) Darcy's law was argued to ignore the inertia of the flow and can possibly miss important characteristics of the flow in the organ: \enquote{given the architecture of a placental circulatory unit, the [maternal blood] jet penetration essential to good utilization of the full villous tree for mass transfer can occur only with appreciable fluid inertia}~\citep{Erian1977}. The validity of the last statement may however depend on the location of the maternal vessels~\citep{chernyavsky_2010}. Our approach operates with an average velocity of the maternal blood in the organ which includes the inertia implicitly;

\item
The advantage of the~PMM is that the results are obtained analytically in a simple form.

\end{enumerate}

\subsection{Other ways of comparison}
Several other ways of validation of the presented model can be suggested, all of them requiring experimental measurements of some quantitative criteria of oxygen transfer efficiency and histomorphometry of placental sections:
\begin{enumerate}
\item 
For each pregnancy, medical doctors possess approximate quantitative criteria of newborns' health, such as the birth-weight to placenta-weight ratio. If such information were obtained together with histological sections of the placenta in each case, a correlation could be studied between the observed villi density, the optimal villi density predicted by the model and the health of the baby. Such experiment, however, requires the development of image analysis techniques that would allow for automatic segmentation and morphometric measurements to be performed on large histological placenta sections;

\item
In the primate placenta (the structure of which is similar to that of the human placenta), more information can probably be obtained from the experiment than it is ethical in the human placenta. Thus, (i)~if one measured the average linear~MBF velocity in the~IVS of a placentone or in the spiral arteries supplying it, and (ii)~if morphometrical measurements were obtained after birth for the same placentone, then the relation between the blood velocity and the observed villi density could be compared with Fig.~\ref{fig:VilliDensityVelocity}. If additionally blood oxygenation were measured in the spiral arteries and the decidual veins, the dependence of oxygen uptake of a placentone on the~MBF velocity could be compared to Fig.~\ref{fig:MaximalUptakeVelocity}. A systematic study of several placentones of the same or different placentas could further improve validation of the model;

\item
Artificial perfusion experiments could also provide data comparable to the results of the model. In such setups, one can estimate the oxygen uptake by the fetal circulation of a given placenta, while having control over perfusion parameters. After the perfusion, histomorphometrical measurements can be performed. The fact that in the artificial perfusion experiments no-hemoglobin blood is normally used does not hamper the comparison, as such situation can be simulated in our model by allowing~$B=1$. Experimental measurements of villi density and of oxygen uptake as a function of the perfusion parameters can then be compared to the predictions of the model.

\end{enumerate}
These suggestions represent possible directions of further development of experimental techniques.

\subsection{Model assumptions and other remarks}

The assumption of fetal villi being a perfect sink erases differences between concurrent and counter-current organizations of flows since this difference essentially arises from the non-uniformity of oxygen concentration in the fetal blood.
The presented STPM can describe oxygen transfer in both these flow orientations under this assumption. As for the cross-current~(multivillous) blood flow organization, the model geometry can be changed to account for the angle~$\alpha$ between the maternal and fetal flows directions.
We have not incorporated such calculations into our model, because~(i) there does not seem to be any fixed co-orientation of the maternal and fetal blood flows in the human placenta, and~(ii) the experimentally observed villi density is already a result of sectioning at an unknown (random) angle to fetal villi. 
If the value of this angle were known, the effect of the sectioning angle could be approximately corrected by multiplying the villi density and the villi perimeter used in the model by the factor~$1/\cos(\alpha)$ before comparison to values observed in experimental slides.

Finally, it should be noted that the curves in Fig.~\ref{fig:UptakeResults} do not reach the expected limit of~1 for the optimal villi density, but stop at~0.75.
This is explained by the circular villi shapes used in the calculations.
The maximal packing density for the used radii ratio of~$\RNum/r\approx14.63$ is known to be~$\phi\approx0.83$ or~$N=177$ cylinders~\citep{Specht2009}. Without using the optimal packing algorithms, we have carried out calculations up to the density~$\phi=0.75$, which we believe to be sufficient to represent densely packed regions of placenta slides~(cf. Figs~\ref{fig:HealthyAndAbnormalPlacentaSections},~\ref{fig:CalculatedGeometries}).
The latter value is the limit of the curves seen in Figs~\ref{fig:UptakeResults}.

\section{Conclusions}

We presented a 2D+1D stream-tube placenta model of oxygen transfer in the human placenta.
Our model incorporates only the most significant geometrical and hemodynamic features, enabling it to remain simple while yielding practical results. 

Modeling oxygen transport in the human placenta traditionally involved the following parameters: incoming~($c_0$) and maximal oxygen concentrations~($\cMax$), diffusivity of oxygen in blood~($D$), solubility of oxygen in blood~($\kHenry$), oxygen-hemoglobin dissociation coefficient~($\bSixty$), density of blood~($\rhoBl$), length of the blood flow path~($L$), a characteristic size of the exchange region~($R$) and~the local permeability of the feto-maternal interface~($w$) or the integral diffusing capacity of the organ~($\DP$). The construction of the STPM showed that several additional parameters are required to account for a~2D placental section structure: (i)~the linear maternal blood flow velocity in the exchange region~($u$, in \unit{m/s}), since the usual integral flow (in \unit{ml/min}) is not enough; (ii)~the volumetric density~($\phi$) of small fetal villi in a given region; and (iii)~the effective radius of fetal villi~($r$), which accounts for the non-circular shape of fetal villi by providing the right volume-to-surface ratio. Although we have indirectly estimated these parameters from the available experimental data, their direct measurement in the human placenta is a promising direction of further development of experimental techniques.

In spite of its simplicity, the model predicts the existence of an optimal villi density for each set of model parameters as a trade-off between the incoming oxygen flux and the absorbing villus surface. The predicted optimal villi density~$0.47\pm0.06$ is compatible with experimentally observed values. 
Dependence of the optimal characteristics on the length of the stream tube and the velocity of the~MBF was investigated. 

In a perspective, one can check the effect of relaxing some of the model assumptions.
One direction of the future work consists in relaxing the slip boundary conditions and comparing the results of the two approaches, which are expected to describe different limiting cases of the real \emph{in vivo} flow.
One can also study the effect of villi shapes and sizes distributions on the predicted optimal villi density and maximal uptake.
Besides, the placenta is a living organ with its own oxygen consumption rate, which may be considered proportional to the weight of its tissues. Once this effect is taken into account, the amount of oxygen transferred at the same villi density from mother to the fetus should decrease, shifting the position of the optimal villi density peak to the left~(Fig.~\ref{fig:UptakeVilliDensityFixedL}). 
Another direction is to apply the STPM to~2D histological slides in order to relate geometrical structure of placenta regions to their oxygen uptake efficiency and hence to be able to distinguish between healthy and pathological placentas. 

\section{Acknowledgements}

This study was funded by the International Relations Department of Ecole Polytechnique as a part of the PhD project of A.S.~Serov, by Placental Analytics~LLC,~NY, by ANR SAMOVAR project n$^\circ$ 2010-BLAN-1119-05 and by ANR project~ANR-13-JSV5-0006-01. The funding sources did not influence directly any stage of the research project.

\section{Conflicts of interest}

The authors declare to have no potential conflicts of interest.

\section{Contributions of authors}

All authors have contributed to all stages of the work except for numerical calculations performed by A.S.~Serov. All authors have approved the final version of the article.

{
\appendix
\renewcommand{\topfraction}{0.45}					

\begin{center}
\Large
Appendices

\end{center}

\renewcommand{\thesection}{\Alph{section}}
\renewcommand\thefigure{{\thesection}.\arabic{figure}}    
\renewcommand{\theequation}{\Alph{section}.\arabic{equation}}
\renewcommand\thetable{\thesection.\arabic{table}}    
\counterwithin{figure}{section}
\counterwithin{table}{section}
\counterwithin{equation}{section}

\section{Parameters of the model}
\label{appendix:sect:ValuesOfModelParameters}

\paragraphBold{Maximal oxygen concentration in the maternal blood~($\cMax$)}
$\cMax$ is the maximal oxygen concentration, which a unit of blood volume would contain at a \unit[100]{\%} saturation only due to binding to hemoglobin molecules. 
To calculate~$\cMax$ we need to multiply the following quantities:
\begin{itemize}
\item
{\tt concentration of hemoglobin in blood.} 
Pregnant women hemoglobin level at term is~$\unit[0.1194\pm0.0007]{g\ Hb/ml\ blood}$~\citep{Wills1947};

\item
{\tt hemoglobin binding capacity.}
Adult person hemoglobin binding capacity is~$\unit[1.37\pm0.02]{ml\ O_2/g\ Hb}$ \citep{Dijkhuizen1977};

\item 
{\tt amount of substance in~$\unit[1]l$ of oxygen.}
As Avogadro's law states, a liter of any gas taken at normal conditions contains~$\unit[1/22.4]{mol}$ of substance.
\end{itemize}

Using these data we obtain
\begin{align*}
\cMax&=0.1194\ \frac{\unit{g\ Hb}}{\unit{ml\ blood}}
\cdot
1.37\ \frac{\unit{ml\ O_2}}{\unit{g\ Hb}}
\cdot
\frac 1 {22.4}\ \frac{\unit{mol}}{\unit{l\ O_2}}
\\
&=
\unit[7.30\pm0.11]{mol/m^3}
\end{align*}

\paragraphBold{Oxygen-hemoglobin dissociation constant~($B$)}

Oxygen-hemoglobin dissociation constant~$B$ represents the interaction of oxygen with hemoglobin in Eq.~\fm{fmGeneralEquationLinear}, when the Hill equation has been linearized. 

Substituting $\cMax=\unit[7.30\pm0.11]{mol/m^3}$, $\bSixty=\unit[0.0170\pm0.0003]{mmHg^{-1}}$ (slope of the linearized Hill equation, see Section~\ref{sectLinearizationOfTheHillsLaw}), $\kHenry\approx\unit[7.5\cdot10^{5}]{mmHg\cdot kg/mol}$~(see Section~\ref{sectInteractionOxygenHemoglobin})
and $\rhoBl\approx\unit[10^3]{kg/m^3}$ (as close to the density of water) into Eq.~\fm{fmGeneralEquationLinear}, we obtain~$B=94\pm2$.

\paragraphBold{Oxygen concentration at the entrance to the~IVS~($c_0$)}

$c_0$ is the concentration of oxygen dissolved in the blood plasma of maternal blood entering the system.
To calculate the concentration~$c_0$, we need to know oxygen saturation of maternal blood which enters the placenta. We will then be able to calculate~$c_0$ by the formula which follows from the Hill equation~\fm{fmSaturationDefinitionHillEquation}:
\begin{linenomath}
\[
c_0=\cPl(z=0)=\frac 1 {B-1}\cdot S(\pONew)\cdot\cMax.
\]
\end{linenomath}

After 12-th week postmenstruation,~$\pONew$ was found to be~\unit[90]{mmHg} in the uterine artery, \unit[61]{mmHg} in the~IVS and~\unit[47]{mmHg} in the decidual vein~\citep{Challier2003,Rodesch1992,Jauniaux2000}.
Taking the IVS value of~\label{parIntervillousPartialPressureValue}\unit[61]{mmHg} (no error estimation available) together with the hemoglobin saturation curve~\citep{Severinghaus1979} we get hemoglobin saturation~$S(\unit[61]{mmHg})\approx\unit[85]{\%}$. Supposing that it is equal to the hemoglobin saturation at the entrance to the~IVS,
we then obtain~$c_0=\unit[(6.7\pm0.2)\cdot10^{-2}]{mol/m^3}$.

\paragraphBold{Oxygen diffusivity in blood plasma~($D$)}

Oxygen diffusivity in the blood plasma has been measured under different conditions:~$D=\unit[(1.7\pm0.5)\cdot 10^{-9}]{m^2/s}$~\citep{Wise1969,Moore2000a}.

\paragraphBold{Effective villi radius~($r$)}

Histological slides of the human placenta show that the cross-sections of fetal villi are not circular. Traditionally, the radius of the villi is estimated by averaging the size of a cross-section in all directions. This yields a villi radius~$\unit[25\text{--}30]{\mu m}$ for terminal and mature intermediate villi, which contribute the most to the uptake~\citep{Sen1979}. However, such calculation ignores the fact that the real perimeter of a villus is larger than that of a circle with the same radius. At the same time, the absorbing surface of villi is important for transport processes as, in the first approximation, oxygen uptake is proportional to its area. 

To let our model appropriately account for both cross-sectional area~($\SSmallVilli$) and the absorbing perimeter~($\PSmallVilli$) of the small villi, an effective villi radius~($r$) is introduced to give the same perimeter-to-cross-sectional-area ratio~($\varepsilon=\PSmallVilli/\SSmallVilli$, also called surface density) as in histological slides. 
In our model,~$\varepsilon=2\pi rN/(\pi r^2N)=2/r$, and we then define the effective villi radius as~$r=2/\varepsilon$ so as to give the right perimeter-to-area ratio for the villi cross-sections. The value of~$\varepsilon$ is taken from histomorphometrical studies.


Experimental values of the perimeter-to-area ratio~$\varepsilon$ and radii calculated from them are shown in Table~\ref{tbl:HistomorphometricEffectiveVillusRadius}. The area~($\SSmallVilli$) and the perimeter~($\PSmallVilli$) of the small villi cross-sections are usually presented in terms of small villi volume~($\vSmallVilli$) and small villi surface area~($\ASmallVilli$) respectively by scaling them to the size of the whole placenta. In spite of this recalculation procedure, the same formula remains valid for~$\varepsilon$:~$\varepsilon=2\ASmallVilli/\vSmallVilli$. One then gets~$r=\unit[41\pm3]{\mu m}$ for the effective villi radius~(a value slightly larger than the average villi radius as it accounts for the correct perimeter-to-area ratio of villi cross-sections).

\begin{table*}[t!]
\newcommand{\tblWidth}{\textwidth}
\caption{
\rm
Estimations of the effective villi radius~$r$. Average small villi fraction in all villi volume~(\unit[66.5]{\%}) and surface area~(\unit[77.3]{\%})~\citep{Sen1979} were used for calculations, when these data were not directly available from the experiment.
}
	\smallskip
	\label{tbl:HistomorphometricEffectiveVillusRadius}
		\begin{tabularx}{\tblWidth}{@{}l*{6}{X}@{}}
		\toprule
		&\citet{Mayhew2000}
		&\citet{Nelson2009}
		&\citet{Aherne1966}
		&\citet{Lee1995}
		&\citet{Mayhew1993Conductance}
		\\
		\midrule
		Small villi volume~($\vSmallVilli$), $\unit{cm^3}$	&	130 & 220 & 190 & 110 & 150
		\\
		Small villi surface area~($\ASmallVilli$), $\unit{m^2}$	& 7 & 11 & 9 & 5 & 7
		\\
		Perimeter-to-area ratio~($\varepsilon$), $\unit[10^{4}]{m^{-1}}$ & 5.4 & 5.0 & 4.7  & 4.5 & 4.7
		\\
		Effective villi radius~($r$), $\unit{\mu m}$ & 37 & 40 & 43 & 44 & 42
		\\
		\bottomrule
		\end{tabularx}\par
\end{table*}

\paragraphBold{Permeability of the effective feto-maternal interface~($w$)}

The permeability of the feto-maternal interface is a local transfer characteristic, which has been studied before under the form of~\textit{oxygen diffusing capacity}~($\DP$). We show below how the two characteristics are related and how one can be recalculated into the other. Following~\citet{Mayhew1984,Bartels1962}, we define~$\DP$ as a volume of gas~(measured at normal conditions) transferred per minute per average oxygen partial pressure difference between the maternal and fetal blood:~%
$\DP=\vOxygen/(t(\overline{\pMat-\pFet}))$,
where the horizontal bar denotes averaging over the feto-maternal interface surface.
Oxygen diffusing capacity has been calculated for the placenta as a whole, but the details of calculations~(namely, rescaling) demonstrate its local nature~\citep{Aherne1966,Laga1973,Teasdale1982,Teasdale1983ClassB,Mayhew1984}.
The following points need to be clarified:
\begin{enumerate}

\item Our model assumption that fetal blood is a perfect sink implies that the partial pressure of oxygen~($\pONew$) appearing in our model is not its absolute value in the maternal blood, but the difference in oxygen partial pressures on both sides of the membrane:~$\pONew=\pMat-\pFet$. This partial pressure can be recalculated into the concentration of oxygen dissolved in the blood plasma~($\cPl$) with Henry's law~(Appendix~\ref{appendix:sect:MathematicalFormualtion}):~$\pONew=\cPl \kHenry/\rhoBl$.
The volume of transferred oxygen~$\vOxygen$ measured at normal conditions can be recalculated into amount of substance with the help of Avogadro's law:~$\vOxygen=\nuOxygen \vMolar$, where~$\vMolar=\unit[22.4]{l/mol}$. One then obtains
\begin{linenomath}
\begin{align}
&\DP=\frac\nuOxygen t\ \frac{\vMolar \rhoBl}{\kHenry\cPlAverage}
&&\text{or}
&&\frac\nuOxygen t=\frac{\DP\kHenry\cPlAverage}{\vMolar \rhoBl},
\label{fm:NuOverT}
\end{align}
\end{linenomath}
where the horizontal bar denotes averaging over the feto-maternal interface surface;

\item
Using the boundary conditions froom Appendix~\ref{appendix:sect:MathematicalFormualtion}, we can write oxygen uptake as~%
$F = -\int_\ASmallVilli D\ \partial\cPl/\partial\nMF\ d S=
w\int_\ASmallVilli c\ dS= w \cPlAverage\ASmallVilli$, %
where~$\ASmallVilli$ is the absorbing surface of fetal villi along the blood flow. The permeability~$w$ can then be rewritten as~$w=F/(\ASmallVilli \cPlAverage)$. By definition, oxygen uptake~$F$ is the amount of oxygen transferred to the fetus~($\nuOxygen$) divided by the time of exchange~$t$, so for~$w$ one gets
\begin{linenomath}
\begin{equation}
w=\frac\nuOxygen t\ \frac1{\cPlAverage \ASmallVilli}.
\label{fm:wFromOxygenUptake}
\end{equation}
\end{linenomath}

\end{enumerate}
Substituting Eq.~\fm{fm:NuOverT} into Eq.~\fm{fm:wFromOxygenUptake}, one relates the permeability~$w$ to the experimental diffusive capacity:
\begin{linenomath}
\begin{equation*}
w= \frac{\DP\kHenry}{\vMolar\rhoBl\ASmallVilli}.
\end{equation*}
\end{linenomath}

The mean value of the diffusing capacity is reported to be~$\DP=\unit[3.9\pm1.0]{{cm}^3/(min\cdot mmHg)}$; as calculated from structural and physical properties of the organ~\citep{Aherne1966,Laga1973,Teasdale1982,Teasdale1983ClassB,Mayhew1984}. The absorbing surface of villi~$\ASmallVilli=\unit[7.8\pm2.3]{m^2}$ is that of small villi (terminal and mature intermediate) as they correspond to the main site of exchange (Table~\ref{tbl:HistomorphometricEffectiveVillusRadius}). %
Using these data, we finally get $w=\unit[(2.8\pm1.1)\cdot 10^{-4}]{m/s}$. Note also that calculations of~\citet{Mayhew1993Conductance} show that oxygen diffusive capacity in diabetic pregnancies, although slightly higher, is of the same order of magnitude as in healthy pregnancies.

\paragraphBold{Radius of the placentone at term~(at least~37 completed weeks of pregnancy)~($R$)}

A normal placenta numbers~$50\pm10$~\citep[p.~161]{benirschke_pathology} placentones, whereas the average radius of the placental disk is~$\unit[11\pm2]{cm}$~\citep{Salafia2012}. Division of the disk area by the number of functional units yields~$R=\unit[1.6\pm0.4]{cm}$. 
A placentone of this radius contains about~$1.5\cdot10^5$ fetal villi of radius~$r$. Numerical resolution of the problem for such a large number of villi would be too time-consuming.
At the same time, oxygen uptake depends rather on villi density than on the number of villi.
We have thus performed the calculations for a smaller radius~$\RNum=\unit[6\cdot10^{-4}]{m}$, and then rescaled oxygen uptake to the real size~$R$ of the placentone by multiplying it by~$R/\RNumSquared$.

Note that the analysis of the characteristic time scales of the transfer processes in the human placenta shows that during the time spent in a placentone by each volume of blood, the characteristic distance of oxygen diffusion in a cross-section is of an~IVS pore size~(about~$\unit[80]{\mu m}$, see Sect.~\ref{sect:TimeScales}). That means that the solution of the diffusion-convection equation in any part of the stream-tube volume that is farther from its boundary than that distance, will be independent of the shape of the boundary, and hence, any shape of the external boundary of a placentone can be chosen.

\paragraphBold{Maternal blood flow velocity~($u$)}

Information on the linear maternal blood flow in the~IVS of the human placenta \emph{in vivo} is scarce. To our knowledge, no direct measurements have ever been conducted. Among the indirect estimations, the following approaches are worth attention:
\begin{enumerate}

\item
The modeling by~\citet{Burton2009} concludes that blood velocity at the ends of spiral arteries falls down to~$\unit[0.1]{m/s}$ due to dilation of spiral arteries towards term.
MBF velocity in the~IVS must drop even further;
\gdef\uFirst{$\unit[0.1]{m/s}$}

\item
The order of magnitude of the MBF velocity can also be estimated as follows: the total incoming flow of blood can be divided by the surface of the whole placenta cross-section occupied by the~IVS. The total incoming flow is known to be of the order of~\unit[600]{ml/min}~\citep{Bartels1962,Browne1953,Assali1953}, or~\unit[450]{ml/min} if one considers that~\unit[25]{\%} of the flow passes through myometrial shunts~\citep{Bartels1962}; cross-section area of the placental disk is~$\pi(\unit[11]{cm})^2\approx\unit[380]{cm^2}$; and the percentage of the cross-section occupied by the~IVS is around~$\unit[35]{\%}$~\citep{Aherne1966,Laga1973,Bacon1986,Mayhew1986}. There is no contradiction between the villi density of around~0.5~(main text, Table~2) and the value~0.35, as the former is the fraction of small villi in a cross-section containing small villi and~IVS only~(all other components subtracted), whereas the latter also accounts for large villi and non-IVS components of a placenta cross-section.  Finally, for the estimation of the average~MBF we get:~$u\sim\unit[450]{ml/min}/(\unit[380]{cm^2}\cdot0.35)\sim\unit[5\cdot10^{-4}]{m/s}$;
\gdef\uSecond{$\unit[5\cdot10^{-4}]{m/s}$}


\item
A more accurate estimation of the blood velocity can be obtained from the results of the experimental study by~\citet{Burchell1967}, in which angiographic photographs of the human uterus at different stages of pregnancy have been obtained by X-rays with a radioactive dye.
The time during which the radioactive dye was observed in the placenta~($\tObs$) and the corresponding diffusion diameter of the dye spot~($2a$) at term were reported~(Table~\ref{tblRadioactiveDiffusionBurchell1967}). The diffusion diameter is defined as the mean maximal diameter of the zone of propagation of the radioactive material, and the dye observation time~($\tObs$) is the time elapsed since the appearance of the radioactive dye till its disappearance from the~IVS. 

\begin{table}
\caption{Radioactive dye diffusion in the human~IVS at term~\citep[original data from][]{Burchell1967}}
\label{tblRadioactiveDiffusionBurchell1967}
\centering
\begin{tabularx}{\columnwidth}{@{}Xl@{}}
	\toprule
	Diffusion diameter~($2a$),~$\unit{cm}$	&	$\leqslant3.5$
	\\
	Dye observation time~($\tObs$),~$\unit{s}$	&	$30$
	\\
	Incoming blood flow per~1 placentone~($q$),~$\unit{ml/min}$	&	$24$
	\\
	Transit time of blood through the~IVS~($\tauPas$),~$\unit{s}$		&	28
	\\
	Mean velocity of blood in the~IVS~($u$),~\unit{m/s}	&	$6\cdot10^{-4}$
	\\
	\bottomrule
\end{tabularx}
\end{table}

Data in Table~\ref{tblRadioactiveDiffusionBurchell1967} shows that the diameter of the diffusion space where the radioactive dye was observed~(up to $\unit[3.5]{cm}$) is of the order of the placentone diameter~($2a\approx\unit[2.5]{cm}$), which justifies the conclusion of the author that in this study, blood propagation in placentones has been observed. The dye volume~$\vDye$ was reported to be~$\unit[50]{cm^3}$ in all the experiments. For estimations, we neglect the effect of its further dissolution and of increase of the dye volume and assume that the dye volume is equally distributed between~$\nCot=50$ placentones of the placenta at term~\citep{benirschke_pathology}. 

From the presented data, one can estimate the total blood flow in one placentone~($q$, in~$\unit{ml/min}$), the IVS blood flow velocity~($u$) and the time~($\tauPas$) that takes a blood volume to pass a placentone. For this purpose, we assume that the~IVS has the shape of a hemisphere of radius~$R$, and blood flows uniformly from its center in radial directions to the hemisphere boundary~\citep[this description is confirmed by the \enquote{doughnut} shape of dye diffusion clouds reported in the study and corresponds to the current views, see][]{chernyavsky_2010,benirschke_pathology}. In this case,~$q=(2\pi a^3/3+\vDye/\nCot)/\tObs$ and 
the transit time of maternal blood through the placentone is~$\tauPas=2\pi a^3/(3q)$. The mean velocity of blood in the placentone can be defined as~$u=a/\tauPas$.

\gdef\uThird{$\unit[6\cdot10^{-4}]{m/s}$}
The results of the calculations of~$q$,~$u$ and~$\tauPas$ are presented in the lower part of Table~\ref{tblRadioactiveDiffusionBurchell1967}. The average~IVS blood flow velocity is estimated to be of the order of~\uThird.

\end{enumerate}

The three estimates~($u_1<$ \uFirst, $u_2\approx$ \uSecond, $u_3\approx$~\uThird) are compatible, and for calculations we use the value~$u=\unit[6\cdot10^{-4}]{m/s}$ obtained from the third, the most precise method. We emphasize that the~MBF velocity may significantly differ between placental regions, and the value~$u=\unit[6\cdot10^{-4}]{m/s}$ should be understood as a mean value for the whole placenta.



\paragraphBold{Cylinder length~($L$)}

{
\tolerance=800 
The length of the cylinder~$L$ is fixed to provide the time of passage~($\tauPas=\unit[27]{s}$) as calculated from~\citet{Burchell1967} (Table~\ref{tblRadioactiveDiffusionBurchell1967}) if the velocity of the flow $u\approx\unit[6\cdot10^{-4}]{m/s}$:~$L\approx\unit[1.6]{cm}$  (no error estimation available). %
Note that this value is comparable to, but smaller than the basic estimate as a doubled placenta thickness (for the idealized case when blood goes from the basal plate to the chorionic plate and then back) because: (i) a placentone may not occupy all the thickness of the placenta; (ii) the central cavity is likely to have less resistance to the flow and larger blood velocities, and should hence probably not be considered as a part of the exchange region; and (iii) blood flow in the exchange region is not directed \enquote{vertically} between the basal and the chorionic plates, but rather from the boundary of the central cavity to decidual veins (see Fig.~\ref{fig:PositioningOfTheGeoemtry}).


}

\section{Mathematical formulation}
\label{appendix:sect:MathematicalFormualtion}

\subsection{Interaction between bound and dissolved oxygen}
\label{sectInteractionOxygenHemoglobin}

The very fast oxygen-hemoglobin reaction can be accounted for by assuming that the concentrations of oxygen dissolved in the blood plasma~($\cPl$) and of oxygen bound to hemoglo\-bin~($\cBnd$) instantaneously mirror each other's changes. Mathematically, both concentrations can be related by equating oxygen partial pressures in these two forms.

Because of a low solubility, the partial pressure of the dissolved oxygen can be related to its concentration by using Henry's law of ideal solution:~$\pONew=\cPl\kHenry/ \rhoBl$, %
where~$\rhoBl\approx\unit[1000]{kg/m^3}$ is the density of blood, and the coefficient~$\kHenry$ can be estimated from the fact that a concentration~$\cPl\approx\unit[0.13]{mol/m^3}$ of dissolved oxygen corresponds to an oxygen content of~$\unit[3]{ml\ O_2/l\ blood}$ or a partial pressure of oxygen of~$\unit[13]{kPa}$ at normal conditions~\citep{Law1999}. No error estimations are provided for these data. These estimates give the Henry's law coefficient:~%
$
\kHenry\sim\unit[7.5\cdot10^{5}]{mmHg\cdot kg/mol}
$
for oxygen dissolved in blood.

The partial pressure of hemoglobin-bound oxygen depends on its concentration through oxygen-hemoglobin dissociation curve~(the Hill equation):
\begin{linenomath}
\begin{align}
\label{fmSaturationDefinitionHillEquation}
&\cBnd=\cMax S(\pONew), &&S(\pONew)\equiv \frac{(\kHill\pONew)^\alpha}{1+(\kHill\pONew)^\alpha},
\end{align}
\end{linenomath}
with~$\kHill\approx\unit[0.04]{mmHg^{-1}}$ and~$\alpha\approx2.65$, that we obtain by fitting the experimental curve of~\citet{Severinghaus1979}, and~$\cMax$ is the oxygen content of maternal blood at full saturation.

Equilibrium relation between~$\cPl$ and~$\cBnd$ can then be obtained by substituting Henry's law into the Hill equation: 
\begin{equation}
\label{fmMirrorRelation}
\cBnd=\cMax\cdot S\left(\frac{\cPl\kHenry}{\rhoBl}\right).
\end{equation}

\subsection{Diffusive-convective transfer}

Diffusive-convective transfer is based on the mass conservation law for the total concentration of oxygen in a volume of blood:
\begin{linenomath}
\begin{equation}
\label{fmMassConservationLaw}
\partialFrac{(\cPl+\cBnd)}{t}+\div \jCTot=0,
\end{equation}
\end{linenomath}
where~$\jCTot=-D\vec\nabla\cPl+\vec u(\cPl+\cBnd)$ is the total flux of oxygen, transferred both by diffusion and convection for the dissolved form and only by convection~(RBCs being too large objects) for the bound form; and %
$\vec u$ denotes the velocity of the~MBF. Omitting the time derivatives in the stationary regime,  substituting the expression for oxygen flux into Eq.~\fm{fmMassConservationLaw} and assuming~$z$ being the direction of the~MBF, we obtain
\begin{linenomath}
\[
\Delta\cPl=\frac uD \partialFrac{(\cPl+\cBnd)}{z}.
\]
\end{linenomath}
Using the relation~\fm{fmMirrorRelation} %
between the dissolved and bound oxygen concentrations, we obtain an equation which contains only oxygen concentration in blood plasma:
\begin{linenomath}
\begin{equation}
\label{fmMainEquationNonLinear}
\Delta\cPl=\frac u D\partialFrac{ }{z}\left(\cPl+\cMax S\left(\frac \kHenry \rhoBl \cPl\right)\right)\!.
\end{equation}
\end{linenomath}
This equation is nonlinear as $\cPl$ appears also as an argument of the Hill saturation function~$S(\pONew)$~(Fig.~\ref{fig:HillsLaw}). 

\global\def\eighthFigureCaption{
Oxygen-hemoglobin dissociation curve. On the figure:~(i) the dots are experimental data at normal conditions as obtained by~\citet{Severinghaus1979}; %
(ii) the curved line shows our fitting of these data by the Hill equation with the coefficients being~$\alpha=2.65$, $k=\unit[0.04]{mmHg^{-1}}$; 
(iii) the straight dashed line is our linear approximation in the~$\unit[{[0,60]}]{mmHg}$ region, slope of the line being~$\bSixty\approx\unit[0.0170\pm0.0003]{mmHg^{-1}}$ (linear fit~$\pm$ standard error).
}
\begin{figure}[tbH!]
\center{\includegraphics[width=3in,
natwidth=1905, natheight=1254]
{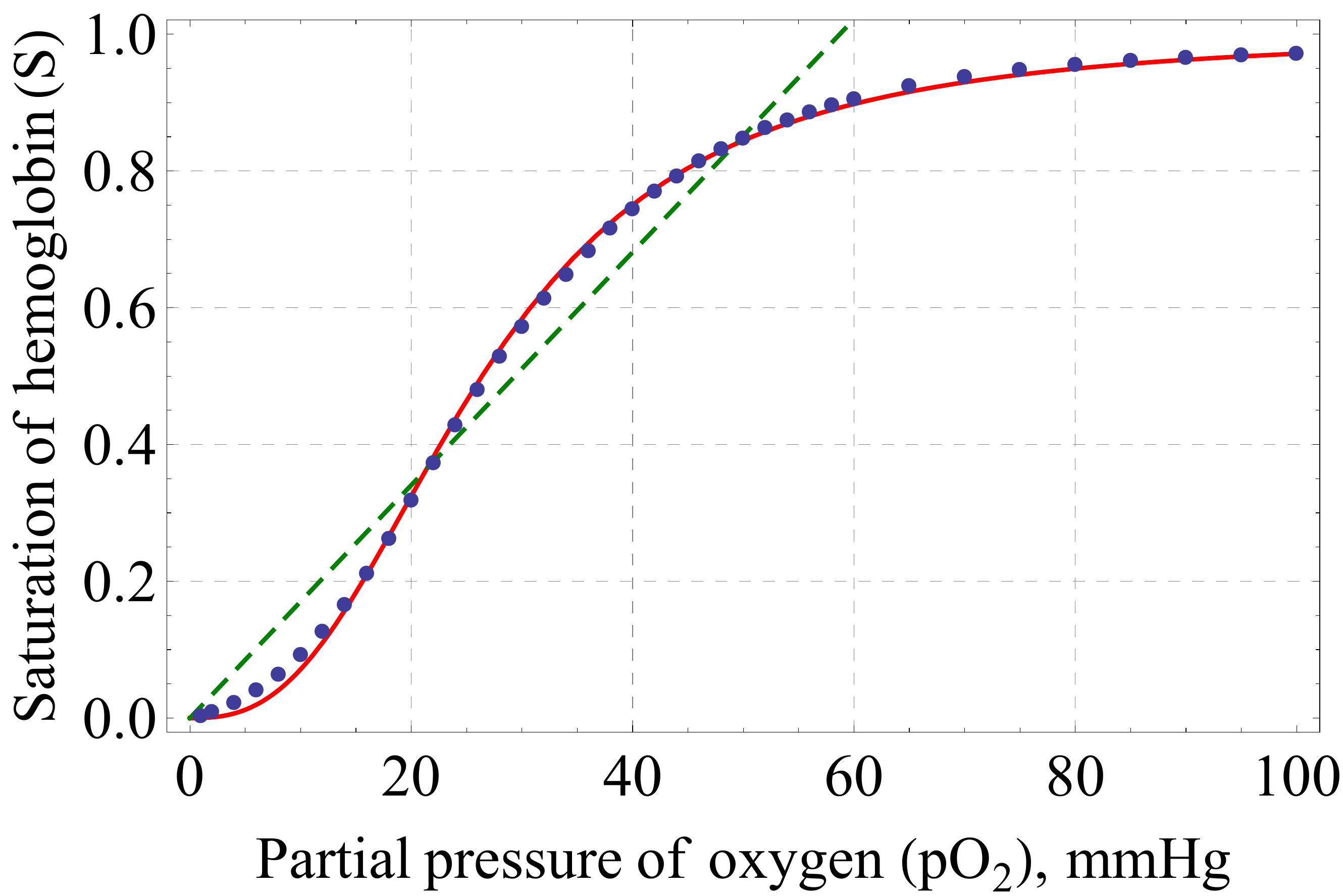}}
\caption{
\eighthFigureCaption
} \label{fig:HillsLaw}
\end{figure}


\subsection{Linearization of the Hill equation}
\label{sectLinearizationOfTheHillsLaw}

Data found in the literature
show that the maternal blood at the entrance to the~IVS of the human placenta has a~$\pONew$ of about~$\unit[60]{mmHg}$~(see calculation of~$c_0$ in Appendix~\ref{appendix:sect:ValuesOfModelParameters}). It is natural to suppose that this pressure is the maximal value in the exchange region. From Fig.~\ref{fig:HillsLaw}, one can see that in the range~$\unit[{[0,60]}]{mmHg}$ the shape of the dissociation curve can be approximated by a straight line. In that case, the~3D equation~\fm{fmMainEquationNonLinear} can be converted to a~2D+1D equation which allows for faster computations, while keeping the main features of the solution. Such linearized equation can be also studied analytically.
Fitting the experimental curve in the region~$\unit[{[0,60]}]{mmHg}$ by a straight line passing through zero we obtain a linear approximation~$S(\pONew)\approx\bSixty\, \pONew, 	\bSixty\approx\unit[0.017]{mmHg^{-1}}$,
with the help of which the diffusion-convection equation~\fm{fmMainEquationNonLinear} transforms into
\begin{linenomath}
\begin{align}
\label{fmGeneralEquationLinear}
&\Delta\cPl=\frac uDB\frac{\partial \cPl}{\partial z}, 
\\
&\text{where } B\equiv\frac\cTot\cPl=1+\frac{\cMax\bSixty\kHenry}{\rhoBl}.
\nonumber
\end{align}
\end{linenomath}
The total oxygen uptake is determined as an integral of the normal component of the total flux of oxygen
\begin{linenomath}
\begin{equation}
\jCTot=-D\vec\nabla \cPl +\vec u\cPl B %
\label{fm:jCTotDefinition}
\end{equation}
\end{linenomath}
across the total surface of the fetal villi. Note that the described linearization procedure is similar to that used for the introduction of an effective oxygen solubility in blood~\citep{Bartels1964,Opitz1952}.

\subsection{Boundary conditions}
\label{sect:boundary_conditions}

Equation~\fm{fmGeneralEquationLinear} should be completed with boundary conditions derived from the model assumptions:
\begin{itemize}
\item there is no uptake at the boundary of the placentone~(of the large external cylinder):~$\partial\cPl/\partial\nMF=0$,
where~$\partial/\partial n$ is a normal derivative directed outside the~IVS;

\item oxygen uptake at the feto-maternal interface is proportional to the concentration of oxygen dissolved in the IVS:~($\partial/\partial\nMF+ w/D)\cPl=0$,
where~$w$ is the permeability of the effective feto-maternal interface;

\item the concentration of oxygen dissolved in the maternal blood plasma is uniform at the entrance of the stream tube~(at~$z=0$ plane):~$\cPl(x,y,z=0)=c_0$, $\forall (x,y)\in \SM$,
where~$\SM$ is the maternal part of a horizontal cross-section.

\end{itemize}

\begin{figure*}[tb]
\setlength{\abovecaptionskip}{15pt}
\centering
\newcommand{\picHeight}{1.25in}
\vspace{-4ex} 
\subfloat[]{
\hspace{-2.4ex}
\includegraphics[height=\picHeight,
natwidth=2720, natheight=2720]
{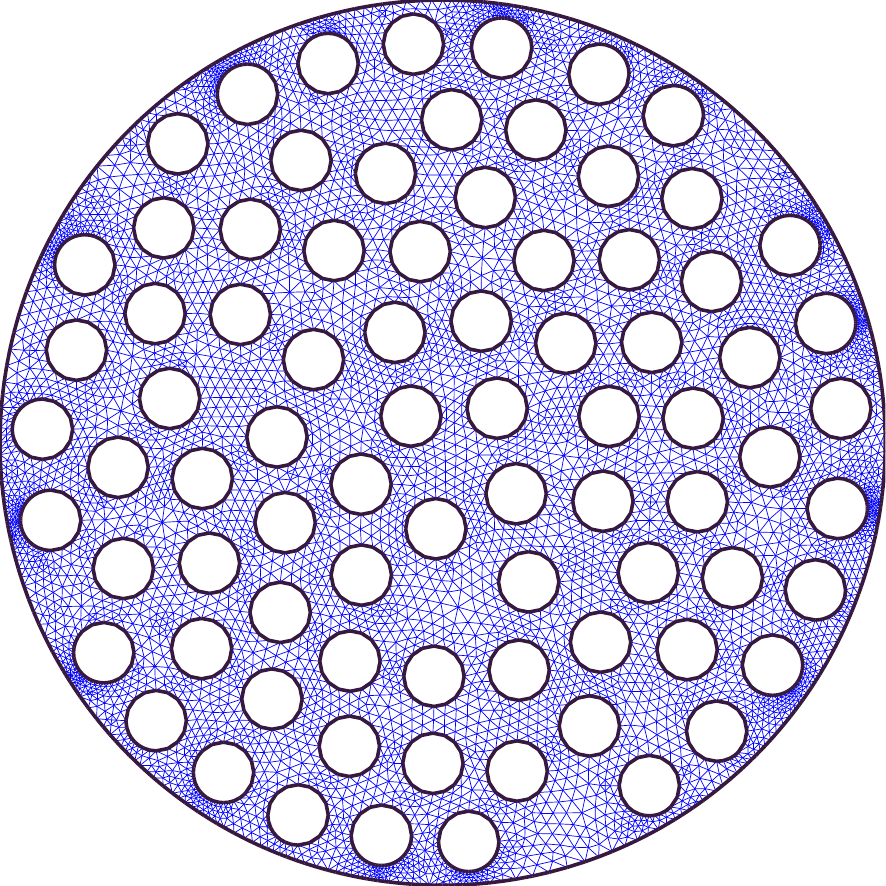}
\label{fig:TypicalMesh}
}
\subfloat[]{
\includegraphics[height=\picHeight, natwidth=1280, natheight=1104]
{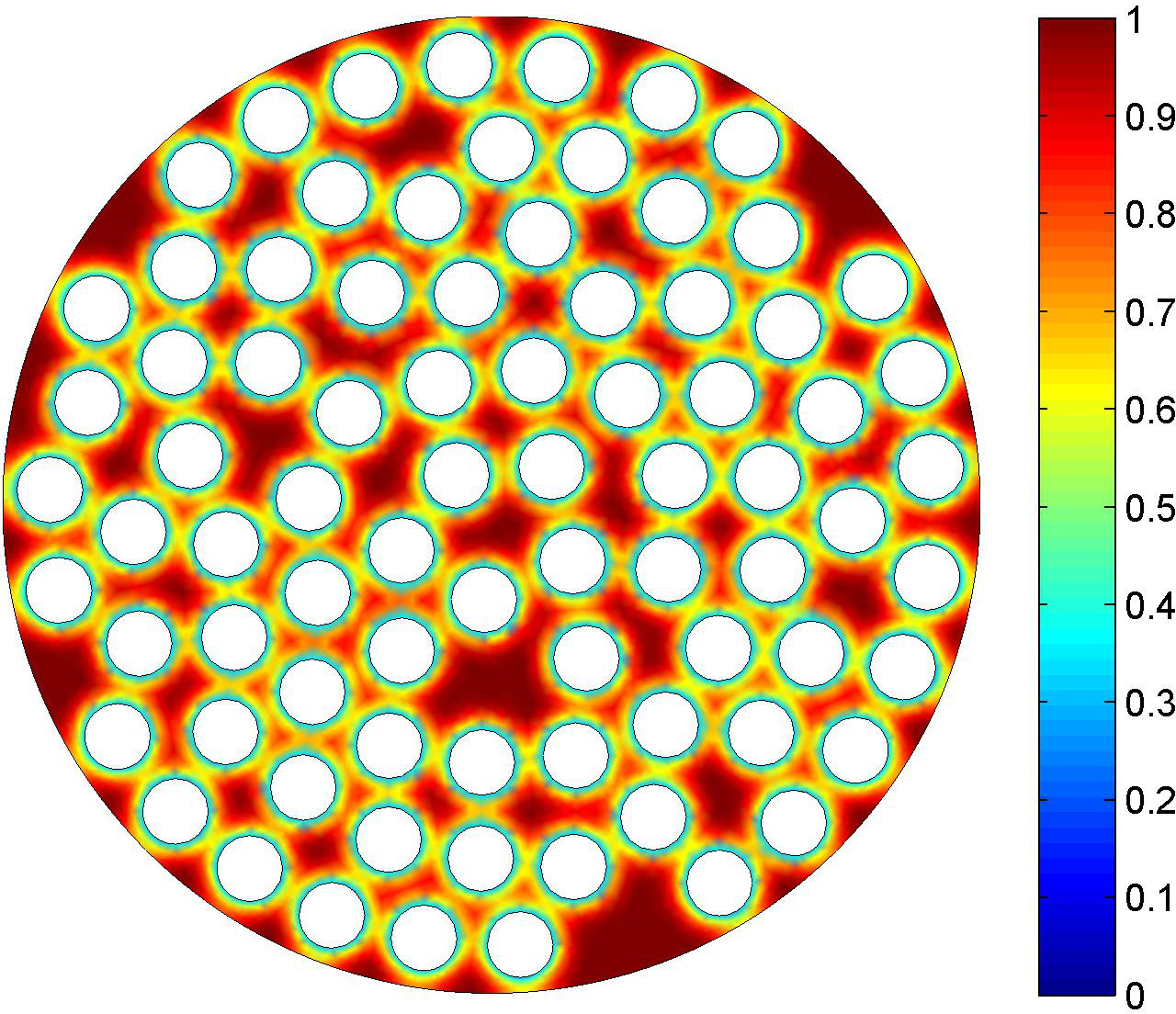}
\label{fig:ConcentrationDistributionLOverThree} }
\vspace{-2ex}
\subfloat[]{
\includegraphics[height=\picHeight, natwidth=1280, natheight=1104]
{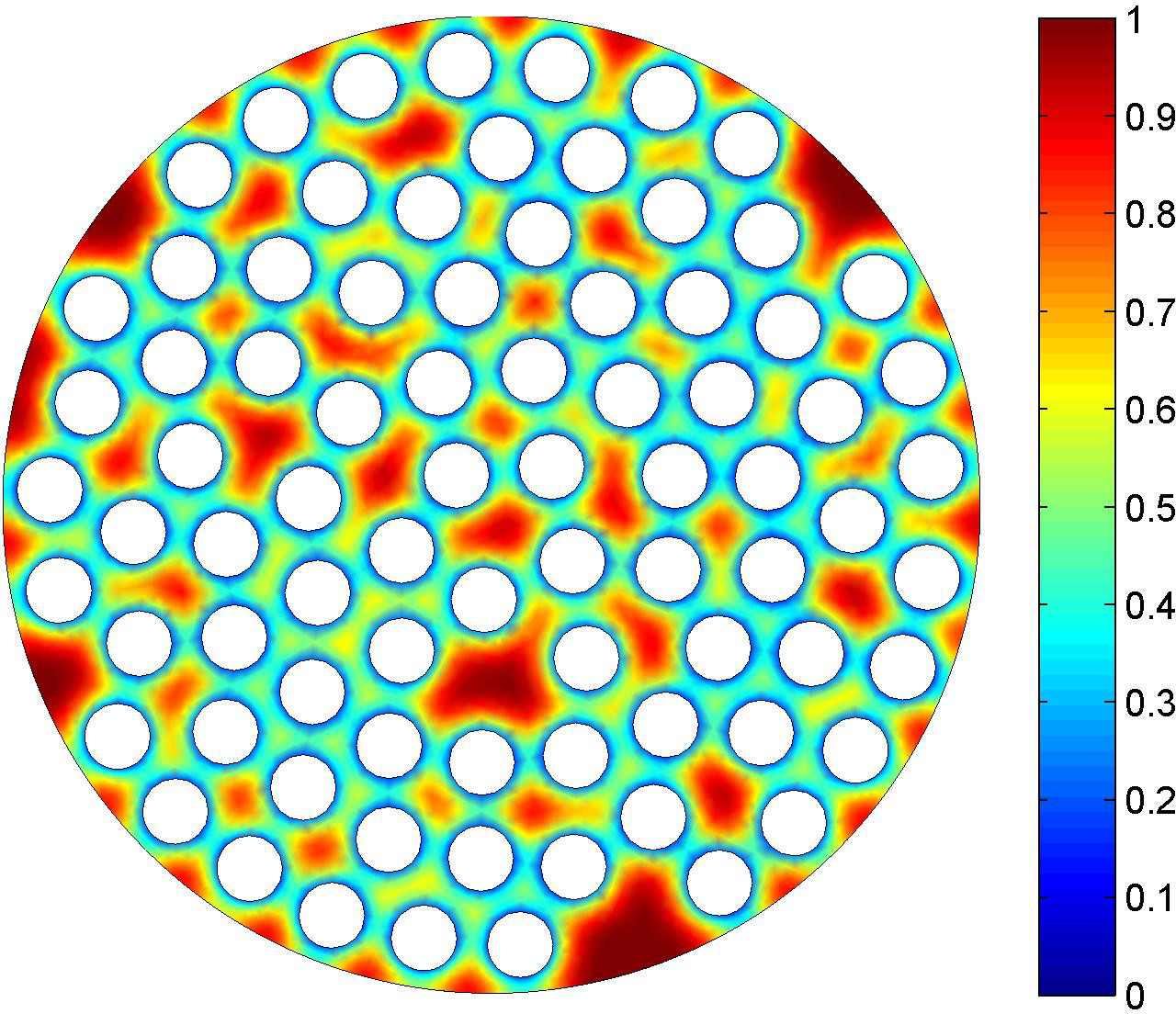}
\label{fig:ConcentrationDistributionTwoLOverThree} }
\subfloat[]{
\includegraphics[height=\picHeight, natwidth=1280, natheight=1104]
{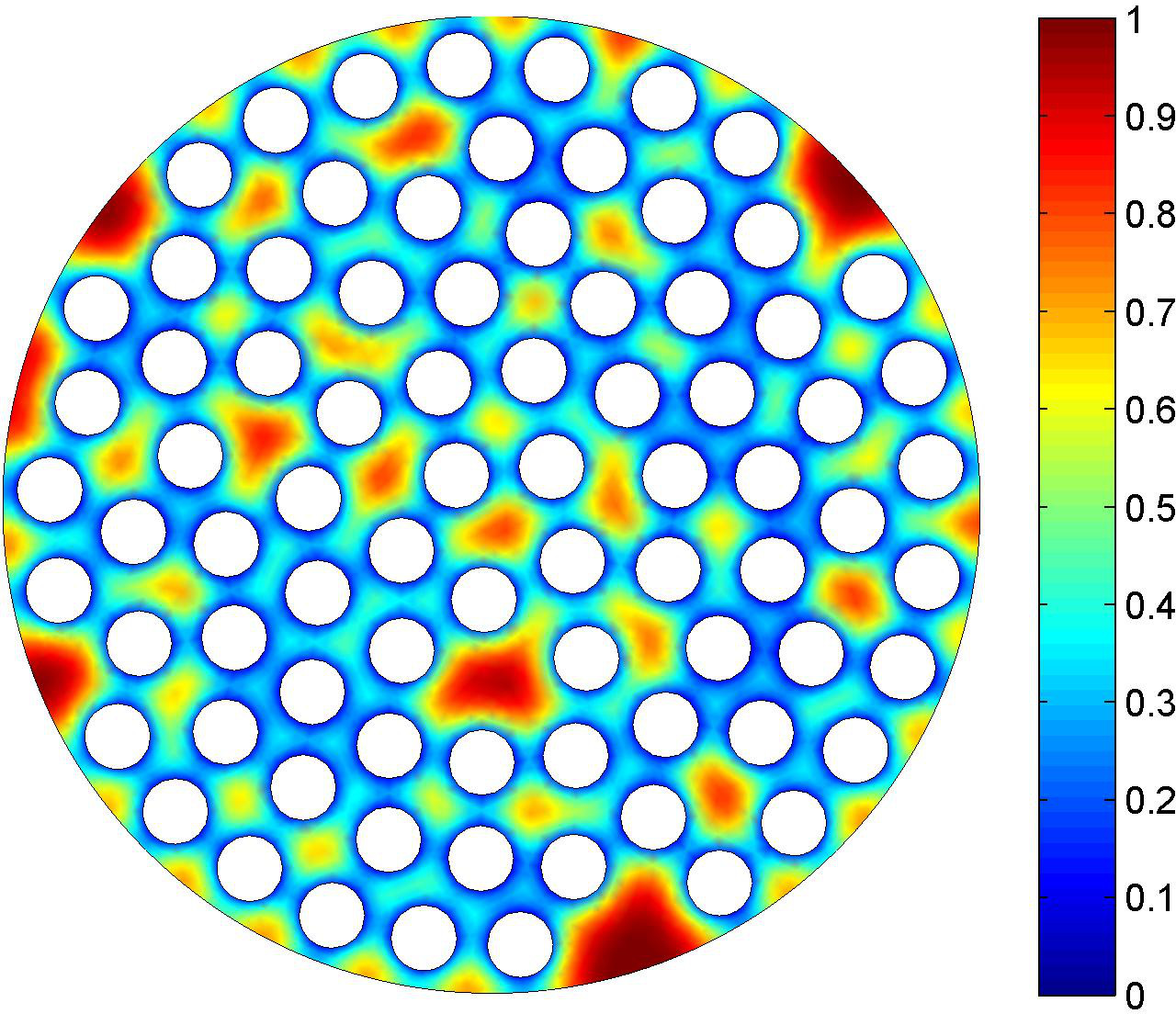}
\label{fig:ConcentrationDistributionL} }
\caption{
\protectedSubref{fig:TypicalMesh} A typical calculation mesh for~$N=80$ villi. Plates~\protectedSubref{fig:ConcentrationDistributionLOverThree}, \protectedSubref{fig:ConcentrationDistributionTwoLOverThree}, \protectedSubref{fig:ConcentrationDistributionL} show oxygen concentration distribution at stream-tube lengths~$z=L/3$, $z=2L/3$ and~$z=L$ respectively for the same geometry. For this illustration, the incoming oxygen concentration is supposed to be~1, and the color axis represents oxygen concentration in the range~$[0,1]$. All the figures are plotted with~4500 eigenfunctions.
}
\label{fig:MeshAndConcetrationIllustration}
\end{figure*}


\subsection{Numerical solution and validation}

Equation~\fm{fmGeneralEquationLinear} is then solved numerically in the PDE~Toolbox\registeredTrademark\ in Matlab\registeredTrademark\ with the help of spectral decomposition over Laplacian eigenfunctions~$\{v_j\}$ in a 2D section~\citep{Davies1996}:
\begin{linenomath}
\begin{align*}
&\cPl=c_0\sum_j a_j v_j \e^{-\mu_j L},
\\ 
&F(L)=c_0\sum_ja_j^2(D\mu_j+uB)(1-\e^{-\mu_j L}),
\end{align*}
\end{linenomath}
where oxygen uptake~$F(L)$ at the stream-tube length~$L$ is calculated from the law of conservation of mass.
The mesh of the~2D geometry of the~IVS necessary for an implementation of a finite elements method is created with Vorono\"i-Delaunay algorithm~(Fig.~\ref{fig:TypicalMesh}). 
Figures~\ref{fig:ConcentrationDistributionLOverThree}--\ref{fig:ConcentrationDistributionL} illustrate oxygen concentration~$\cPl$ distribution in a~2D cross-section calculated at three different stream-tube lengths.

Convergence of the results has been tested by varying the number of eigenfunctions found and the number of mesh points. Figure~\ref{fig:SolutionConvergence} shows the dependence of the uptake at length~$L$ on the number of eigenfunctions. 
One can see that a good convergence is achieved with around~3000 eigenfunctions for all geometries and 
for all three lengths shown in the article~($z=L/3$, $z=2L/3$ and~$z=L$).

\begin{figure*}[t]
\centering
\newcommand{\picWidth}{3.25in}
\subfloat[]{
\hspace{-2.4ex}
\includegraphics[width=\picWidth,
natwidth=2720, natheight=2720]
{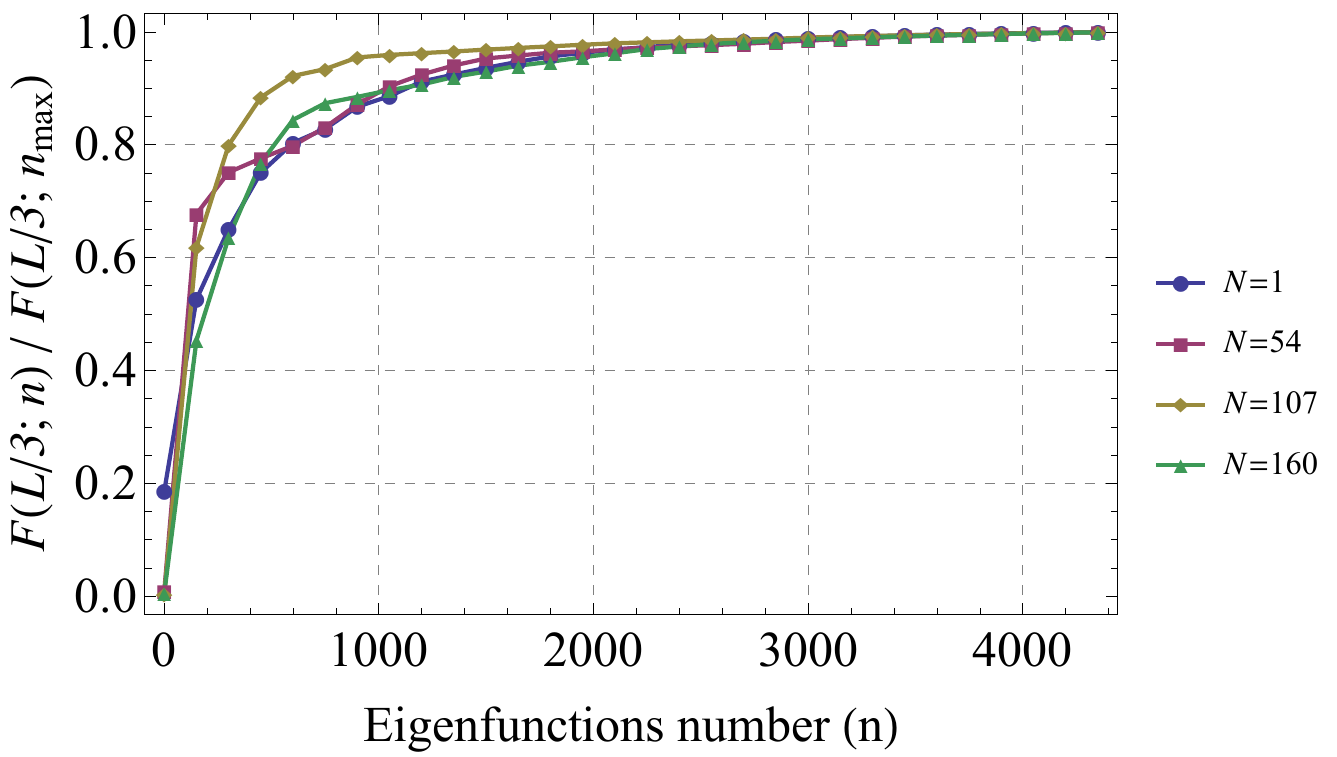}
\label{fig:SolutionConvergenceLOverThree}
}
\subfloat[]{
\includegraphics[width=\picWidth,
natwidth=2720, natheight=2720]
{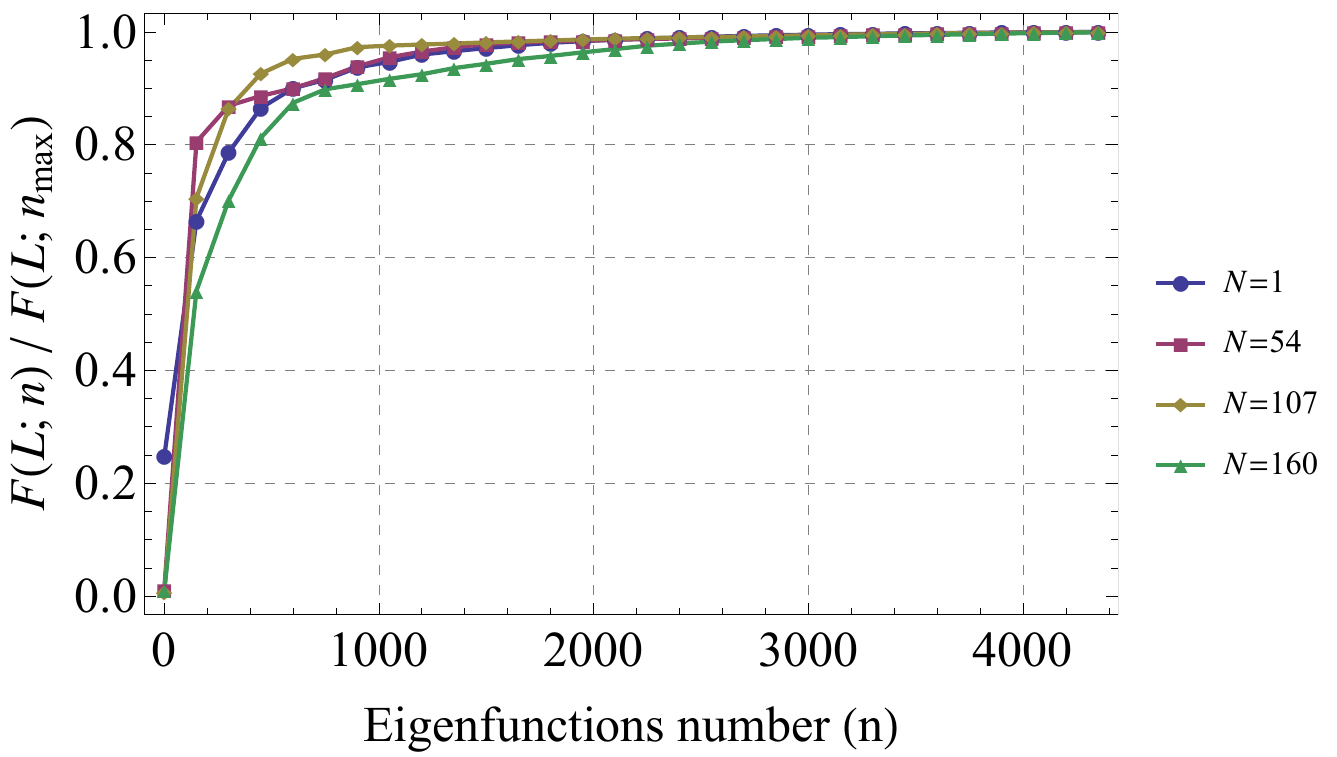}
\label{fig:SolutionConvergenceL}
}
\caption{
Convergence of oxygen uptake~$F(z;n)$ at stream-tube lengths~$z=L/3$~\protectedSubref{fig:SolutionConvergenceLOverThree} and~$z=L$~\protectedSubref{fig:SolutionConvergenceL} with the number of eigenfunctions~$n$ for different numbers of villi~$N$. The uptake for each geometry is normalized to the value~$F(z;\nMax)$ obtained for the maximal calculated number of eigenvalues~$\nMax=4500$. One can see that the convergence is slower at smaller lengths.
} 
\label{fig:SolutionConvergence}
\end{figure*}


The concentration~$\cPl(r)$ and oxygen uptake~$F(L)$ for the case of one fetal villus have been compared to the exact analytical solution~\citep{Lundberg1963}:
\begin{linenomath}
\begin{align*}
&v_j(r)=A_j \left(J_0\left(\sqrt{\Lambda_j}r\right)-\frac{J_1\left(\sqrt{\Lambda_j}R\right)}{Y_1\left(\sqrt{\Lambda_j}R\right)}Y_0\left(\sqrt{\Lambda_j}r\right)\right),
\\
&\mu_j=\frac{\sqrt{B^2u^2+4D^2\Lambda_j}-Bu}{2D},
\\
&A_j%
=\left(2\pi\int_{r_0}^R\left(J_0\left(\sqrt{\Lambda_j}r\right)\right.\right.
\\
&\quad\left.\left.-\frac{J_1\left(\sqrt{\Lambda_j}R\right)}{Y_1\left(\sqrt{\Lambda_j}R\right)}Y_0\left(\sqrt{\Lambda_j}r\right)\right)^2rdr\right)^{-1/2},
\\
&a_j=2\pi \int_{r_0}^R A_j \left(J_0\left(\sqrt{\Lambda_j}r\right)-\frac{J_1\left(\sqrt{\Lambda_j}R\right)}{Y_1\left(\sqrt{\Lambda_j}R\right)}Y_0\left(\sqrt{\Lambda_j}r\right)\right)
rdr,
\end{align*}
\end{linenomath}
and $\{\Lambda_j\}$ are defined as roots of the equation
\begin{linenomath}
\begin{multline*}
\left(\frac{w}{D}J_0\left(\sqrt{\Lambda_j}r_0\right)
+\sqrt{\Lambda_j}J_1\left(\sqrt{\Lambda_j}r_0\right)\right)Y_1\left(\sqrt{\Lambda_j}R\right)
\\
-
\left(\frac{w}{D}Y_0\left(\sqrt{\Lambda_j}r_0\right)+\sqrt{\Lambda_j}Y_1\left(\sqrt{\Lambda_j}r_0\right)\right)J_1\left(\sqrt{\Lambda_j}R\right)=0,
\end{multline*}
\end{linenomath}
$J_m(r)$ and~$Y_m(r)$ are the $m$-th order Bessel functions of the first and second kind respectively,~$r_0$ and~$R$ are the radii of the small and large cylinders respectively and~$r$ is the radial coordinate. 
Figure~\ref{fig:ComparisonAnalyticalOneCylinder} shows that the relative error between the numerical and analytical oxygen uptake for the case of one villus is less than~\unit[5]{\%} for all physiologically relevant stream-tube lengths. 


\begin{figure}[tb]
\center{\includegraphics[width=0.37\paperwidth,
natwidth=2720, natheight=2720]
{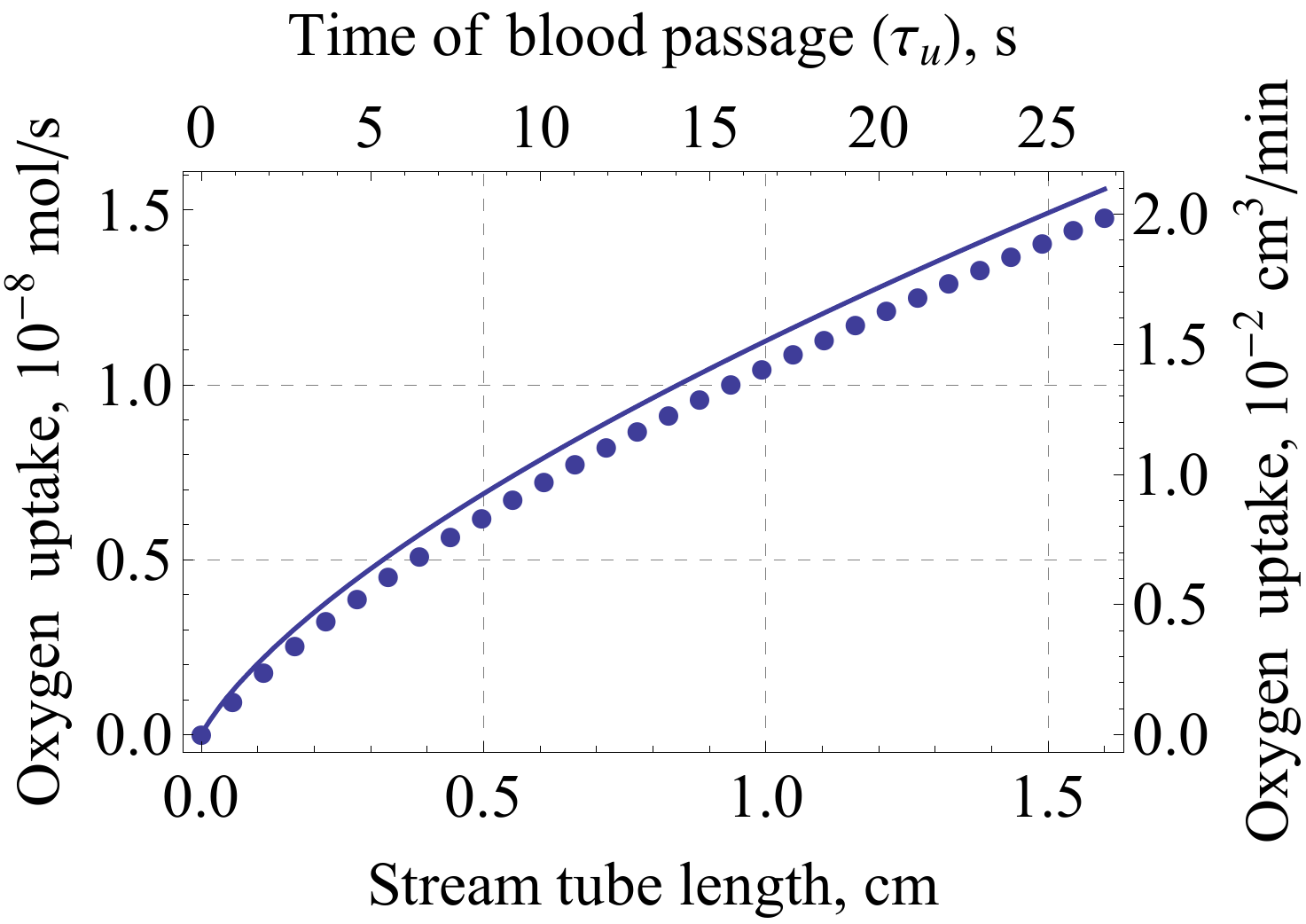}}
\caption{
Comparison of oxygen uptake for one villus calculated numerically with~4500 geometrical eigenvalues~(points) and analytically~(solid line).
} \label{fig:ComparisonAnalyticalOneCylinder}
\end{figure}

\section{Error estimation}
\label{appendix:error_estimation}

The uncertainties of measurement of the parameters of the model can be used to estimate the error in determination of the optimal villi density~$\phi_0$ and the maximal oxygen uptake~$\FMax$. 
If a variable~$f$ depends on several independent variables~$x_i$ with known standard deviations~$\delta x_i$, it is known that its standard deviation~$\delta f$ has the following form~\citep{Taylor1996}:
\begin{linenomath}
\[
(\delta f)^2=\sum_i \left(\frac{\partial f}{\partial x_i} \delta x_i \right)^2.
\]
\end{linenomath}
This formula can be applied to estimate the standard deviations of~$\phi_0$ and~$\FMax$ due to uncertainties of measurement of the parameters of the model, for which error estimations are known~($B$, $D$, $r$, $R$, $w$, $c_0$). Variations of the parameters~$\cMax$, $\bSixty$, $\kHenry$ are included in the variation of~$B$. For the parameters~$u$ and~$L$ no error estimations are available. Variations~$\delta\phi_0$ and~$\delta\FMax$ are then:
\begin{linenomath}
\begin{align*}
(\delta \phi_0)^2&=
\left(\partialFrac{\phi_0}{B} \delta B \right)^2
+
\left(\partialFrac{\phi_0}{D} \delta D \right)^2
+
\left(\partialFrac{\phi_0}{r} \delta r \right)^2
\\
&+
\left(\partialFrac{\phi_0}{R} \delta R \right)^2
+
\left(\partialFrac{\phi_0}{w} \delta w \right)^2
+
\left(\partialFrac{\phi_0}{c_0} \delta c_0 \right)^2,
\nonumber
\end{align*}
\[
\begin{aligned}
(\delta \FMax)^2&=
\left(\partialFrac{\FMax}{B} \delta B \right)^2
+
\left(\partialFrac{\FMax}{D} \delta D \right)^2
+
\left(\partialFrac{\FMax}{r} \delta r \right)^2
\\
&+
\left(\partialFrac{\FMax}{R} \delta R \right)^2
+
\left(\partialFrac{\FMax}{w} \delta w \right)^2
+
\left(\partialFrac{\FMax}{c_0} \delta c_0 \right)^2.
\end{aligned}
\]
\end{linenomath}

Since analytical formulas for the dependence of~$\phi_0$ and~$\FMax$ on the parameters of the model are not available, the partial derivatives appearing in the variations~$\delta\phi_0$ and~$\delta\FMax$ were
calculated numerically. For each variable independently, small deviations were assumed and estimations for the partial derivatives were obtained: 
\begin{itemize}
\item
for~$\phi_0$: 
$\partialFrac{\phi_0}{B}=\unit[-1.6\cdot10^{-3}]{}$, %
$\partialFrac{\phi_0}{D}\approx\unit[0]{s/m^2}$, %
$\partialFrac{\phi_0}{r}=\unit[-3800]{m^{-1}}$, %
$\partialFrac{\phi_0}{R}\approx\unit[0]{m^{-1}}$, %
$\partialFrac{\phi_0}{w}=\unit[550]{s/m}$, %
$\partialFrac{\phi_0}{c_0}\approx\unit[0]{m^3/mol}$; %
\item
for~$\FMax$: %
$\partialFrac{\FMax}{B}=\unit[6.8\cdot10^{-9}]{mol/s}$, %
$\partialFrac{\FMax}{D}\approx\unit[0]{mol/m^2}$, %
$\partialFrac{\FMax}{r}=\unit[-1.2\cdot10^{-2}]{mol/(m\cdot s)}$, %
$\partialFrac{\FMax}{R}=\unit[1.4\cdot10^{-4}]{mol/(m\cdot s)}$, %
$\partialFrac{\FMax}{w}=\unit[1.7\cdot10^{-3}]{mol/m}$, %
$\partialFrac{\FMax}{c_0}=\unit[1.7\cdot10^{-5}]{m^3/s}$. %
\end{itemize}
Note that at the numerical precision,~$\phi_0$ and~$\FMax$ appear to be independent of the diffusivity~$D$ of oxygen. Such behavior cannot be universal for all values of parameters (since~$D=0$, for example, would obviously give no oxygen uptake), but seems to be approximately valid for the normal placenta parameters~(Table~\ref{tblCalculationsParameters}). Substituting the standard deviation of the parameters: %
$\delta B=2$, %
$\delta r=\unit[3\cdot10^{-6}]{m}$, %
$\delta R=\unit[4\cdot10^{-3}]{m}$, %
$\delta w=\unit[1.1\cdot10^{-4}]{m/s}$, %
$\delta c_0=\unit[2\cdot10^{-3}]{mol/m^3}$ (Table~\ref{tblCalculationsParameters}), %
we obtain %
$\delta\phi_0=0.06$ and~$\delta\FMax=\unit[6\cdot10^{-7}]{mol/s}$.


}

\bibliographystyle{my_plainnat}

\small
\setlength{\bibsep}{0pt}
\bibliography{\detokenize{library}}

\end{document}